\documentclass{aa}
\usepackage[super]{nth}
\usepackage{graphicx}
\usepackage{txfonts}
\usepackage{lscape}
\usepackage{float}
\usepackage{soul}

\usepackage{xcolor}
\usepackage{natbib}
\usepackage[pdfpagelabels=false]{hyperref}	
\hypersetup{colorlinks=true,linkcolor=blue,citecolor=blue,filecolor=blue,urlcolor=blue,}

\makeatletter
\renewcommand*\aa@pageof{, page \thepage{} of \pageref*{LastPage}}
\makeatother

\bibpunct{(}{)}{;}{a}{}{,} 

\usepackage{tikz}
\definecolor{lime}{HTML}{A6CE39}
\DeclareRobustCommand{\orcidicon}{
	\begin{tikzpicture}
	\draw[lime, fill=lime] (0,0) 
	circle [radius=0.16] 
	node[white] {{\fontfamily{qag}\selectfont \tiny ID}};
	\draw[white, fill=white] (-0.0625,0.095) 
	circle [radius=0.007];
	\end{tikzpicture}
	\hspace{-2mm}}

\foreach \x in {A, ..., Z}{\expandafter\xdef\csname orcid\x
\endcsname{\noexpand\href{https://orcid.org/\csname orcidauthor\x\endcsname}{\noexpand\orcidicon}}}

\newcommand{\vsini}{\mbox{$\varv\sin i$}}

\newcommand{\vrad}{RV}
\newcommand{\dvrad}{RV$_{\rm pp}$}

\newcommand{\Teff}{\mbox{$T_{\rm eff}$}}
\newcommand{\logg}{\mbox{$\log g$}}

\newcommand{\vcrit}{$v_{\rm crit}$}
\newcommand{\logLs}{$\log (\mathcal{L}/\mathcal{L}_{\odot})$}
\newcommand{\logL}{$\log (L/L_{\odot})$}
\newcommand{\Ls}{$\mathcal{L}$}

\newcommand{\fwhb}{\textit{FW3414(H$\beta$)}}

\let\oldAA\AA
\renewcommand*{\AA}{\,\oldAA\xspace}
\newcommand{\kms}{\,\mbox{km\,s$^{-1}$}\xspace}
\newcommand{\MSol}{\,\mbox{M$_\odot$}\xspace}

\newcommand{\ls}{\mbox{$\lesssim$}\,}
\newcommand{\gs}{\mbox{$\gtrsim$}\,}

\begin{document}
\title{The IACOB project}
\subtitle{XIV. New clues on the location of the TAMS in the massive star domain}
\titlerunning{Main sequence extension of massive stars}
\author{A. de~Burgos\inst{1,2\orcidA{}}, S. Simón-Díaz\inst{2,3\orcidB{}}, M.~A. Urbaneja\inst{4\orcidC{}}, G. Holgado\inst{2,3\orcidD{}},\\ S. Ekström\inst{5\orcidE{}}, M.~C. Ramírez-Tannus\inst{6\orcidF{}}, E. Zari\inst{7\orcidG{}}}
\authorrunning{A. de~Burgos et al.}
\institute{
European Southern Observatory, Alonso de Córdova 3107, Vitacura, Santiago, Chile
\and
Instituto de Astrof\'isica de Canarias, Avenida V\'ia L\'actea, E-38205 La Laguna, Tenerife, Spain
\and
Universidad de La Laguna, Dpto. Astrof\'isica, E-38206 La Laguna, Tenerife, Spain
\and
Universität Innsbruck, Institut für Astro- und Teilchenphysik, Technikerstr. 25/8, A-6020 Innsbruck, Austria
\and
Département d’Astronomie, Université de Genève, Chemin Pegasi 51, 1290 Versoix, Switzerland
\and
Max-Planck Institut für Astronomie (MPIA), Königstuhl 17, 69117 Heidelberg, Germany
\and
Dipartimento di Fisica e Astronomia,Università di Firenze, Via G. Sansone 1, I-50019, Sesto F.no (Firenze), Italy 
}
\date{Received 1 December 2024 / Accepted 27 January 2025}
\abstract 
{Massive stars play a very important role in many astrophysical fields. Despite their scarcity with respect to less-massive counterparts, their influence on the chemo-dynamical evolution of the galaxies is substantial. Yet, some fundamental aspects of their evolution remain poorly constrained. 
In this regard, there is an open debate on the width of the main-sequence (MS) phase, in which stars spend most of their lifetimes.}
{We aim to create an updated Hertzsprung-Russell (HR) diagram that includes a volume-limited and statistically significant sample of massive stars with luminosities $L$\,\gs$2\times 10^{4}L_{\odot}$ and effective temperatures \Teff\,\gs14\,kK. Our goal is to use this sample to investigate the extension and end of the MS while also incorporating information about projected rotational velocities (\vsini) and spectroscopic binarity status.}
{We combined spectroscopic parameters derived with the \textsc{FASTWIND} stellar atmosphere code and \textit{Gaia} distances to obtain estimates of stellar parameters for a sample of 876 Galactic luminous O- and B-type stars gathered within the IACOB project. We used the ALS~III catalog of Galactic OB stars to select the best-suited volume-limited sample for our study. We chose the {\tt iacob-broad} tool to derive \vsini\ estimates and reviewed multi-epoch spectra to identify single- and double-line spectroscopic binaries (SB1, SB2).}
{We present an HR diagram for a sample of 670 stars located within 2500\,pc that has the best balance between completeness and number. We evaluated the extension of the MS in terms of the drop in the relative number of stars as a function of the effective temperature for different luminosity ranges. We found a consistent cool boundary at $\approx$22.5\,kK within the full range of luminosities that we used to delineate the terminal-age main sequence (TAMS).
We obtained a smooth decrease of the highest observed \vsini\ with \Teff\ along the MS band, which is likely limited by the critical velocity. We consider this effect combined with a lower expected fraction of stars beyond the MS as the best explanation for the lack of fast-rotating objects in the post-MS region. Our results favor low to mild initial rotation ($v_{\rm ini}$\,\ls150\kms) for the full sample along with a binary past for the well-known tail of fast-rotating stars in the \vsini\ distribution. The prominence of SB1 and SB2 systems within the MS band and the 25\% decrease in the relative fraction of SB1 systems when crossing the TAMS can further delineate its location.}
{}
\keywords{Stars: massive -- stars: evolution -- stars: rotation -- (Stars:) Hertzsprung-Russell -- (Stars:) binaries: spectroscopic -- techniques: radial velocities} 
\maketitle


\section{Introduction}
\label{section:1_tmp}

Massive stars (M\,\gs8\MSol) represent the minority of stars in the local Universe \citep{salpeter55, chabrier05}. Despite their low relative number, their influence on the interstellar medium and the chemical evolution of their host galaxies is enormous \citep[e.g.,][]{maeder00, langer12, nomoto13, smith14}. Furthermore, they are among the most luminous stars in the Universe, making them useful tools as distance indicators \citep{kudritzki99, kudritzki03b}, and the main stellar contributors to the brightness of high-redshift galaxies \citep[e.g.,][]{dijkstra07}. Despite their influence and importance for many branches of astrophysics, the physical properties and processes that govern their evolution are not yet fully understood. 

As for less-massive objects, massive stars spend most of their lifetimes in the main sequence (MS), a phase in which stars burn their core hydrogen while slowly becoming cooler and larger in size. Upon termination of the MS, stars undergo a rapid reconfiguration of their cores, which translates into an important reduction of the star's surface temperature and a significant increase in their sizes. In the classical Hertzsprung-Russell (HR) diagram, the lack of massive stars with intermediate effective temperatures (between those characterizing the stars on the MS and the red supergiant phase, respectively) caused by this quick transition is known as the Hertzsprung gap. 

Traditionally, blue supergiants (BSGs; B-type stars with luminosity classes I and II) were considered to be post-MS objects following MS O-type stars. However, the discovery of an overdensity of BSGs by \citet[][]{fitzpatrick90} within the Hertzsprung gap raised questions about the true extension of the main sequence of massive stars and the true nature of BSGs.
One possibility that would solve the still persistent uncertainty is considering that a significant fraction of BSGs are MS objects. This would imply a reconsideration of the MS extension in the evolutionary models. 
In this regard, the length of the MS depends not only on the stellar mass and the initial rotation rates of the stars \citep{meynet05, heger05, georgy13} but also on different proposed internal mixing processes that can provide additional hydrogen to the star's core \citep{maeder81, maeder09, schootemeijer19, martinet21}, thus extending the MS.
Furthermore, massive stars are affected by strong stellar winds, which can also affect their lifetimes through the effect of mass loss \citep{kudritzki00, vink10, smith14}.
Another possibility is that additional evolutionary channels contribute to the observed overdensity of BSGs. These may include stars evolving blueward, such as those that undergo a blue loop after the red supergiant (RSg) phase \citep{stothers75, ekstrom12}, but also stars that have evolved through binary interactions \citep{sana12, demink13, menon24}.
In this entangled context, observational constraints are the key to improving the situation.

Since the work by \citet{blaha89} and \citet{fitzpatrick90} using photometric data of stars in the Galaxy and the Large Magellanic Cloud, only the study conducted by \citet{castro14} using a statistically significant sample of massive stars with derived spectroscopic parameters from the literature has attempted to better constrain the width of the MS. Although their sample comprised $\approx$400 Galactic OB stars and their results evidenced the extension of the MS, the presence of observational biases and the lack of homogeneity in their analysis prevented them from quantitatively providing the location of the terminal-age MS (TAMS) for masses greater than 20\MSol. Consequently, high-quality statistically significant samples of BSGs homogeneously analyzed and for which additional spectroscopic information is available (including spin rates, binary status, and surface abundances) are key to improving the current situation, which is the main goal of this work. Following \citet{martinet21}, we define the TAMS by the position in the MS band where the evolutionary tracks reach their lowest effective temperature before momentarily returning to a hotter region (the so-called hook) at core hydrogen exhaustion \citep[see also][]{castro14}.

In this paper, we benefit from the spectroscopic observations gathered within the IACOB project \citep[][]{simon-diaz20} and from the astrometric distances delivered by the \textit{Gaia} mission \citep{gaiacollaboration16, gaiacollaboration22, babusiaux23} to investigate three empirical aspects that can shed light on the extension of the MS and the location of the TAMS in the 12\,--\,40\MSol range of initial masses. These are the distribution of stars in the HR diagram in terms of density and relative number, the spin-rate properties, and the binary status.

This paper is organized as follows. Section~\ref{section:2_tmp} presents the spectroscopic sample of stars and data used to carry out this work. Section~\ref{section:3_tmp} describes the methodology used to evaluate the completeness of our sample. In Sect.~\ref{section:4_tmp} we present the HR diagram of the stars, their empirical properties evaluated in this work, our proposed TAMS compared to the most-used evolutionary models, and a comparison with the results of \citet{castro14}. In Sect.~\ref{section:5_tmp} we discuss the location of the TAMS in the HR diagram and the rotational properties of the sample, as both are connected with further implications for massive star evolution. Section~\ref{section:6_tmp} presents the main conclusions and the follow-up work.


\section{Sample and data}
\label{section:2_tmp}


\subsection{Spectroscopic sample}
\label{subsection:21.spec_sample}

The spectroscopic sample of this work comprises 876 O3- to B6-type stars that cover the ranges of 49\,--\,14\,kK in effective temperature (\Teff), 4.2\,--\,1.8\,dex in surface gravity (\logg), and 6.0\,--\,4.3\,dex in luminosity (\logL).
Of them, 678 were selected from the sample of late O- and early B-type stars presented in \citet{deburgos23a} as stars putatively descended from stellar objects born as O-type stars (in the context of single-star evolution). This sample was later enlarged with additional observations of stars of similar types to increase its completeness (see Sect.~\ref{section:3_tmp}).
The remaining 198 objects were selected from \citet{holgado18, holgado20, holgado22}.
In total, $\approx$45\% of the sample corresponds to stars classified as B-type supergiants and bright giants and $\approx$35\% to O-type stars. The remaining $\approx$20\% corresponds to B-type stars classified as giants to dwarfs.

All stars in the sample have magnitudes of $B_{\rm mag}\ls11$, and their optical spectra were collected from the IACOB spectroscopic database and the ESO public archive. In particular, they were acquired using either FIES \citep[][2.5\,m NOT]{telting14}, HERMES \citep[][1.2\,m Mercator]{raskin11}, or FEROS \citep[][2.2\,m MPG/ESO]{kaufer97} high resolution (R\,=\,$25\,000-85\,000$) optical (3800\,--\,7000\AA) echelle spectrographs. The average signal-to-noise ratio is $\approx$130 \citep[measured at 4500\,\AA; for more details, see][]{deburgos23a}. 

Our sample includes likely single stars and single-line spectroscopic binaries (SB1) in which only one set of spectral lines is detected and where the Doppler shifts indicate the presence of a companion. The sample excludes hypergiants, classical Be-type stars or stars showing strong evidence of being surrounded by a disk, and double- or higher-order spectroscopic binaries (SB2+) where the visual inspection of the spectra shows two or more separated components associated with two or more companions. As also indicated in \citet{deBurgos24a}, the exclusion of these objects is due to the inherent restrictions imposed by the models employed in the quantitative spectroscopic analysis (see Sect.~\ref{subsection:23.spec_param}).


\subsection{Astrometric distances}
\label{subsection:22.distances}

We adopted the distances quoted in \citet{deburgos23a} or obtained them using the same approach. In general, most of them were taken directly from the geometric distances quoted by \citet{bailer-jones21} or, for the few brightest objects, from the parallaxes ($\varpi$) provided by \citet{vanLeeuwen07} using the \textit{Hipparcos} data \citep{perryman97}. The former also includes a direction-dependent prior and a 3D extinction map of the Galaxy that are used in the determination of the distances. On average, our adopted distances have an uncertainty of $\approx$150\,pc.


\subsection{Spectroscopic parameters}
\label{subsection:23.spec_param}

We applied the same methodology as in \citet{deBurgos24a} to derive the spectroscopic parameters for the O9\,--\,B6 stars not analyzed there that are included in the present sample.
Briefly, the line-broadening analysis was performed using the {\tt iacob-broad} tool \citep[see][]{simon-diaz14a}, which allows one to obtain estimates of the projected rotational velocity (\vsini). For the remaining spectroscopic parameters used here, namely \Teff\ and \logg, the analysis was carried out using a grid of unclumped model atmospheres (that is, not considering wind inhomogeneities) computed with the non-local thermodynamic equilibrium model atmosphere and the Fast Analysis of STellar atmospheres with WINDs line synthesis code \citep[\textsc{FASTWIND} v10.4.7,][]{santolaya-rey97, puls05, rivero-gonzalez11, puls20}. Although we refer to Sect.~3 in \citet{deBurgos24a} for additional details on the properties of the grid, the main advantage of our analysis comes from the use of a statistical emulator of synthetic spectra, which is combined with a Markov chain Monte Carlo (MCMC) method to obtain the probability distribution function for the parameters mentioned above. 
For the 198 O-type stars from \citet{holgado18, holgado20, holgado22}, \vsini\ was also obtained using the {\tt iacob-broad}, while \Teff\ and \logg\ were obtained using the {\tt iacob-gbat}/\textsc{FASTWIND} tool \citep{simon-diaz11b, sabin-sanjulian13}. We refer to those works for more details.


\subsection{Fundamental parameters}
\label{subsection:24.fund_param}

Reliable distances fulfilling $\sigma_{\varpi}$/$\varpi$\,\ls0.15 in the full sample \citep[$>$97\% of the total; see][]{bailer-jones21} were used to derive stellar radii and luminosities for the 876 stars.
To do this, we first downloaded the optical $B$ and $V$ magnitudes available from \citet{mermilliod06} and the infrared $J$, $H$, and $Ks$ magnitudes from \citet[][\textit{2MASS}]{skrutskie06}. Then, a synthetic spectral energy distribution (SED) corresponding to the solution obtained in the spectroscopic analysis was calculated in \textsc{FASTWIND}. Following an MCMC approach in which $E(B-V)$ and $R_\mathrm{v}$ were varied, the observed and synthetic SEDs of the star were compared, allowing us to obtain the total V-band extinction ($A_{\rm V}$) with an average uncertainty of 0.07\,mag. 
Combining $A_{\rm V}$ and the apparent $V$ magnitudes with the distances mentioned above, we computed the absolute $V$ magnitudes. Then, following Eq.~1 in \citet{herrero92}, we obtained the stellar radii. In a last step, we used the Stefan-Boltzman law ($L = 4\pi R^{2}\sigma T_{\rm eff}^{4}$) to obtain the luminosities, for which we achieved an average propagated uncertainty of 0.1\,dex.


\subsection{Radial velocity measurements}
\label{subsection:25.multi-epoch}

We obtained radial velocity (\vrad) estimates and peak-to-peak amplitude values (\dvrad) in 605 stars for which multi-epoch spectra in the IACOB spectroscopic database are available. For most O-type stars, we adopted the values from \citet{holgado18}. The methodology followed there consisted of a cross-correlation between the observed spectra and the best-fitting synthetic \textsc{FASTWIND} spectra for a given set of diagnostic lines. For most O9- and all B-type stars in the sample, our estimates were obtained following the methodology described in \citet{deburgos20}. In this case, individual lists of several diagnostic lines are selected and optimized for each sub-spectral type. Then, a combination of line-fitting and sigma-clipping routines allowed us to obtain an average estimate for each spectrum. Alternatively, and only for stars with \vsini\,\gs150\kms, the estimates were derived using a cross-correlation technique similar to \citet{holgado18}.
Peak-to-peak amplitudes were obtained simply as the difference between the largest difference of all the individual values.


\section{Completeness analysis}
\label{section:3_tmp}

\begin{figure}[!t]
\centering
\includegraphics[width=0.45\textwidth]{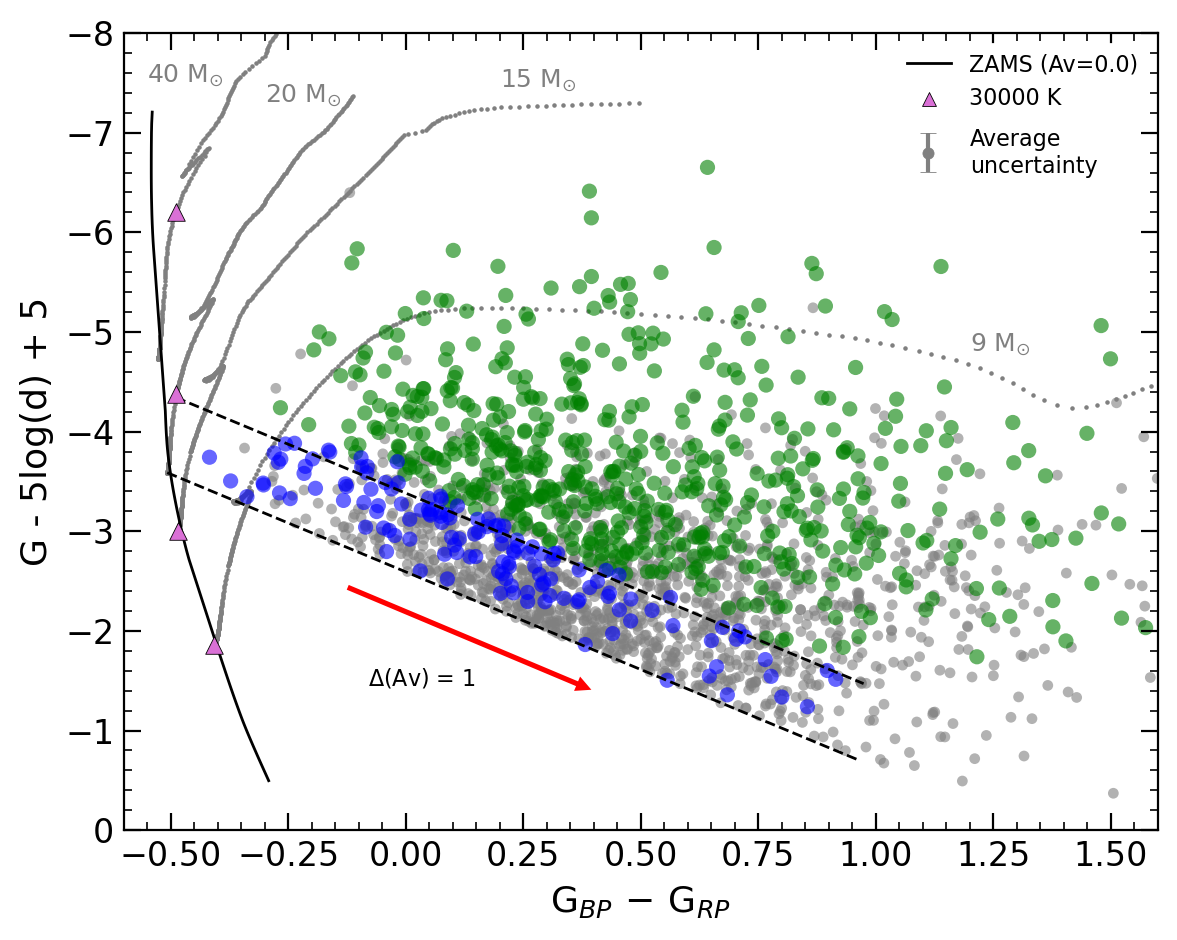}
\caption{\textit{Gaia} CMD including the stars in the sample and other missing stars included in the ALS~III catalog (Pantaleoni González, et al., in prep.). All the stars have $B_{\rm mag}$\,<\,11 and distances below 4\,kpc. The two diagonal dashed black lines mark the reddening line of a 20\MSol\ star at the zero-age MS (ZAMS, bottom) and at 30\,kK (top).
Both are extended up to a $\Delta$(A$_{v}$) = 2. Stars shown in green include those for which both the use of the maximum and minimum distances result in the star above the top reddening line. Those for which minimum distance locate them below the same reddening line are indicated in blue.
Stars missing from the ALS~III catalog are shown in gray. Four evolutionary tracks corresponding to 9, 15, 20, and 40\MSol are included for reference. They were downloaded from the MESA Isochrones \& Stellar Tracks online tool \citep[MIST; see][for references purposes]{dotter16, choi16a, choi16b, paxton13, paxton15} for solar metallicity, no initial rotation, and A$_{V}$ = 0.0. The average uncertainty in the y-axis is indicated in the legend, while it is negligible in the x-axis.} 
\label{fig:gaia_alsIII}
\end{figure}

\begin{table}[!t]
    \centering
    \caption{Completeness summary for our stellar sample with derived stellar parameters with respect to the ALS~III catalog, for stars within same distance and brightness (see main text for further details).}
    \label{tab:completeness}
    \addtolength{\tabcolsep}{-0.2em}
    \begin{tabular}{rcccccc}
        \hline\hline\noalign{\smallskip}
             & \multicolumn{2}{c}{$B_{\rm mag}\leq$9} & \multicolumn{2}{c}{$B_{\rm mag}\leq$10} & \multicolumn{2}{c}{$B_{\rm mag}\leq$11} \\
    Distance &  Total\,\#& \% obs. & Total\,\#  & \% obs.   & Total\,\# & \% obs.  \\
        \hline\noalign{\smallskip}
          $d$\,<1\,kpc &  33 & 97\% &  34 & 94\% &  35 & 94\% \\\noalign{\smallskip}
          $d$\,<2\,kpc & 219 & 99\% & 294 & 90\% & 363 & 77\% \\\noalign{\smallskip}
          $d$\,<3\,kpc & 388 & 99\% & 662 & 76\% & 866 & 62\% \\\noalign{\smallskip}
          $d$\,<4\,kpc & 401 & 98\% & 750 & 71\% &1047 & 54\% \\
        \hline
    \end{tabular}
    \tablefoot{We include the statistics for stars above the top reddening line of Fig.~\ref{fig:gaia_alsIII} (area with green circles) and up to four different distances. Specifically, the different columns indicate the total number of stars with $B_{\rm mag}$ up to 9, 10, and 11, and the associated percentage of observed stars relative to the total.}
\end{table}

\begin{figure*}[t!]
\centering
\includegraphics[width=0.995\textwidth]{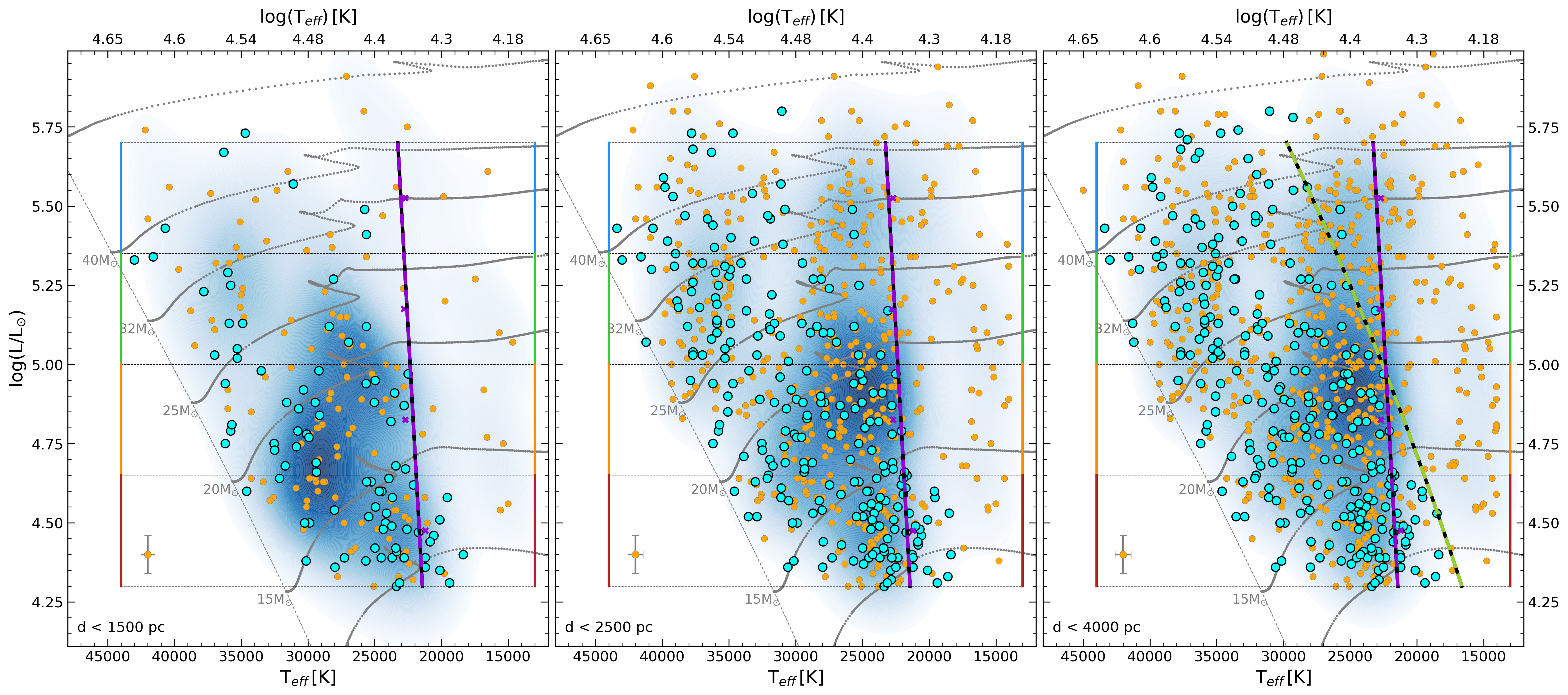}
\caption{Hertzsprung-Russell diagrams showing the stars in our sample. From left to right, each panel limits the sample to stars within 1500 (221 stars), 2500 (670 stars), and 4000\,pc (852 stars) of distance, respectively. The stars with \vsini\,>\,100\kms are indicated with cyan circles, whereas the rest are shown in orange. The average uncertainties are indicated with error bars in the lower-left corner of the panels. Each panel displays four luminosity ranges included for easier interpretation of the other figures. Our proposed location for the TAMS is marked with a purple line in all panels (see Fig.~\ref{fig:vsini_teff} and Sect.~\ref{subsection:41_drop}). The evolutionary tracks are taken from the MIST web-tool for solar metallicity and no initial rotation. The dashed green line on the right panel corresponds to a second-order polynomial fit of the points in the tracks where \vcrit\ reaches 220\kms. A contour density mesh based on a Gaussian kernel is included to represent the density of points.}
\label{fig:main_hrd}
\end{figure*}

In \citet{deburgos23a}, a preliminary analysis of the completeness of the available spectroscopic sample was presented, including a first analysis of the number of observed and missing stars to be included in follow-up observing campaigns (see Table~1 of that work). The main goal there was to corroborate that the sample had a statistically significant number of stars up to a certain distance from the solar neighborhood. In \citet{deBurgos24a}, the study focused on presenting the methodology and estimates of some of the spectroscopic parameters of a sample of $\approx$500 stars, but did not include any analysis of the completeness of that sample. This work includes $\approx$200 additional stars analyzed using the same methodology plus $\approx$200 stars from \citet{holgado18, holgado20, holgado22}, and we evaluated our completeness against the same reference sample: the Alma Luminous Star catalog \citep[][ALS~III; Pantaleoni González et al., in prep.]{pantaleoni-gonzalez21}. 

Figure~\ref{fig:gaia_alsIII} shows our analyzed sample (green and blue circles) along with missing objects included in the ALS~III catalog (gray circles) on a \textit{Gaia} color-magnitude diagram (CMD). The top dashed black line starting at the 20\MSol\ evolutionary track and extending several magnitudes of extinction separates our stars at the approximate age at which O-type stars evolve into B-type stars. As shown in \citet{deburgos23a}, all stars above this line should, in principle, be O-type stars or their descendants when evolving as predicted by single-star evolutionary models, independently of extinction.

For our completeness evaluation, we selected only those stars in the ALS~III with B6 spectral types or earlier and with a $B_{\rm mag}$\,<\,11, as quoted in the SIMBAD astronomical database \citep{weis20}.
This was made to limit the catalog to stars with the same spectral types and magnitude range as in our sample (see also below).
We also removed hypergiant, Be, and SB2+ stars identified as such in previous works, as they are not in our sample (see Sect.~\ref{subsection:21.spec_sample}). Additionally, we removed missing stars identified as classical Be-stars in the BeSS \citep{neiner11} and BeSOS \citep{arcos21} databases, plus some other stars identified as such in SIMBAD.
Last, we removed stars with a distance larger than 4000\,pc in \citet{bailer-jones21} or the ALS~III catalog and those with a poor distance determination, which mostly corresponds to very bright stars with a large parallax uncertainty ($\sigma_{\varpi}$/$\varpi$\,\gs0.15).

Table~\ref{tab:completeness} summarizes the completeness of our sample relative to the ALS~III catalog. It includes information on the number of stars and the observed percentages relative to the total for four different distances up to 4\,kpc. In addition, stars are separated by their $B_{\rm mag}$ into three groups. 
The percentage of analyzed stars with $B_{\rm mag}$\,<\,9 consistently reaches $\approx$100\% throughout all distances considered.
For stars with $B_{\rm mag}$\,<\,10, the percentage decreases from $\approx$90\% to $\approx$70\% at 4\,kpc (below 1\,kpc, the total number is not relevant due to the lack of faint stars).
Here, we must also take into account that, despite the high level of completeness indicated by the ALS~III authors for stars up to 5\,kpc and $B_{\rm mag}$\,$\ls$\,16, the further the distance, the higher the possibility of highly reddened stars along the line of sight not included in the catalog.
In this regard \citet{deburgos23a} showed that up to 2\,kpc, sources within the original sample of O9\,--\,B9 type stars could reach up to 8\,--\,9 magnitudes of extinction and would still have $B_{\rm mag}$\,<\,16, reassuring a good level of completeness in the ALS~III up to that distance.

Based on these results, we concluded that the completeness level achieved relative to the ALS~III catalog is particularly good for stars with $B_{\rm mag}$\,<\,9, and it is sufficiently high up to 3\,kpc (>60\%) to study evolutionary aspects of massive stars, such as the extension of the MS. However, we considered a more conservative distance of 2.5\,kpc to evaluate this and other aspects. In particular, this distance provides a confident 70\% completeness for all stars brighter than $B_{\rm mag}$\,=\,11. 
We also note that the percentages of stars analyzed are approximately similar for all the spectral types and for both hemispheres, minimizing the possibility of biases in that regard (see Appendix~\ref{apen.complet_sptNS}).


\section{Results}
\label{section:4_tmp}


\subsection{The Hertzsprung-Russell diagram: Drop in density}
\label{subsection:41_drop}

\begin{figure*}[!t]
 \centering
    \includegraphics[width=0.7\textwidth]{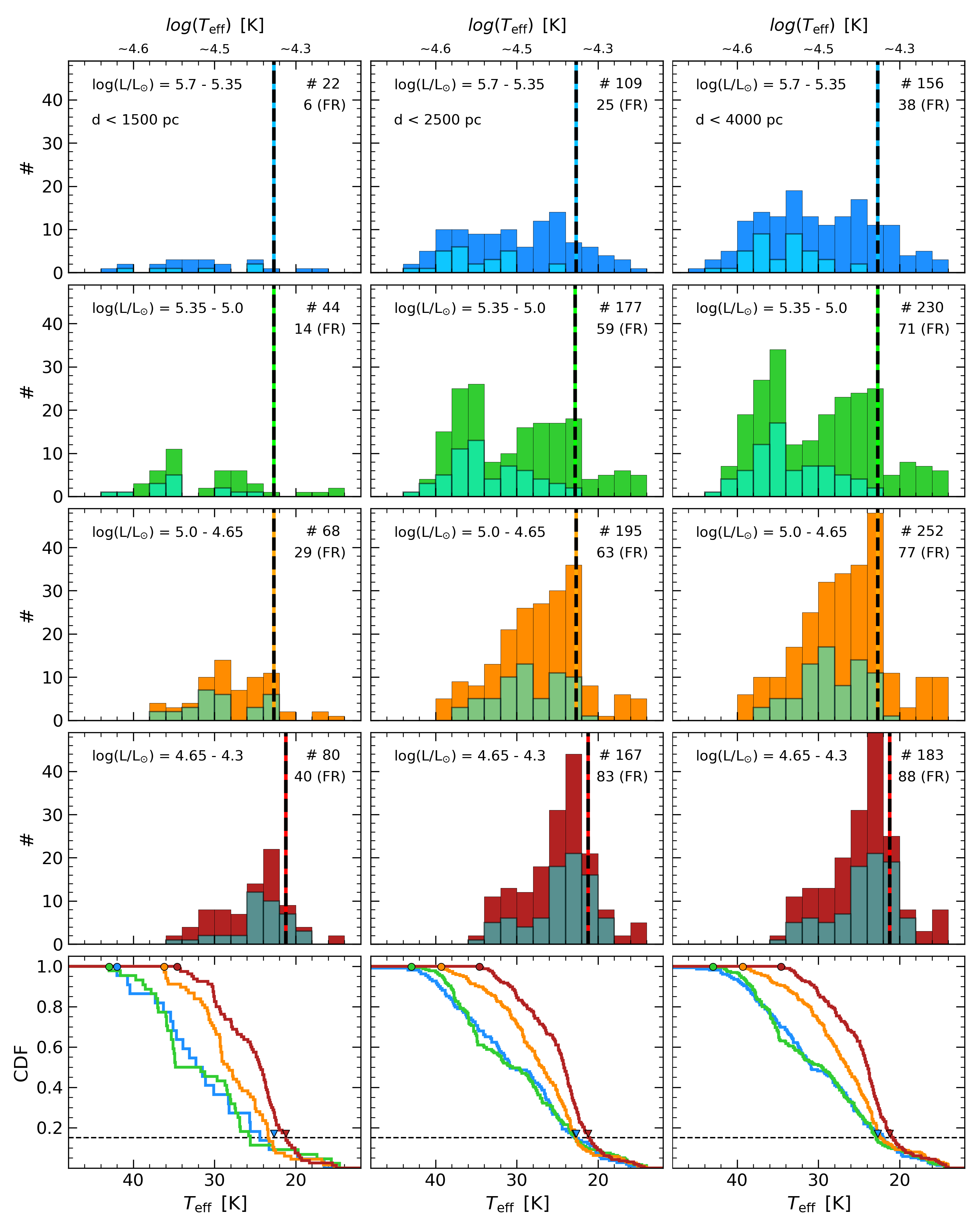}
    \caption{Number of stars with respect to their effective temperature. The histograms in the top four rows of subpanels separate stars within the indicated range of luminosities using different colors. Cyan bins indicate fast-rotating stars with \vsini\,>\,100\kms. Each histogram displays the total number of stars and of fast-rotating objects (FR). Each column separate stars by their distances, as indicated in the top panel. The adopted position of the TAMS for each luminosity range is included with a vertical dashed line (see Sect.~\ref{subsection:41_drop}). The bottom row of subpanels displays the corresponding cumulative distribution function following the color code of the histograms.}
\label{fig:vsini_teff}
\end{figure*}

One of the most robust empirical evidence for the location of the TAMS is the expected drop in the density of stars that populate the HR diagram as a function of \Teff. As explained in Sect.~\ref{section:1_tmp}, this is a consequence of the rapid evolution of massive stars once they reach the end of the MS, the hydrogen is exhausted in their cores, and core nuclear burning stops. In this section, we investigate the occurrence of this feature.

Figure~\ref{fig:main_hrd} includes three different HR diagrams showing our sample of O3\,--\,B6 spectral type stars for which we account with reliable fundamental parameters. Each one considers stars up to a different distance. The central panel shows stars located up to 2500\,pc, that is, to our considered best distance in terms of completeness, and includes a total of 670 stars. The right panel extends the distance to 4000\,pc and displays 852 of the 876 stars of the complete sample. The left panel is limited to stars up to 1500\,pc and includes 221 stars.

Regardless of the distance, the distribution of the stars in the three panels shows an apparent lower density of objects below 23\,--\,21\,kK. To better characterize this feature, Fig.~\ref{fig:vsini_teff} shows the number of stars against their effective temperature on a grid of histograms that separate stars in different distances and luminosity ranges as indicated in the panels (the latter ranges are also included in Fig.~\ref{fig:main_hrd}).
Looking at the histograms for stars within 2500\,pc (middle column), those panels evidence a consistent drop in the relative number of stars below 22\,kK, being less steep only in the top panel of stars in the highest luminosity range. As is discussed in Sect.~\ref{section:5_tmp} (see also Sect.~\ref{subsection:43_stell-ev}), we interpret the drop in density as a consequence of stars leaving the MS and therefore outlining the location of the TAMS.

We used the associated cumulative distribution function (CDF) of the central bottom panel of Fig.~\ref{fig:vsini_teff} to establish a uniform criterion for the position of the drop in \Teff. In particular, we took into account the increase in the CDFs at $\approx$0.15 with increasing \Teff, which is particularly steep for stars with \logL\,<\,5\,dex, where also the number of objects is larger. We decided to adopt this value and the associated \Teff\ in each luminosity range to define the drop.
These four values of \Teff\ are marked with a dashed vertical line in the central histograms and are also copied into the other histograms with stars up to 1500 and 4000\,pc.
We paired each \Teff\ with the average \logL\ of each range to analytically describe the TAMS using a linear fit, which is shown in the three HR diagrams of Fig.~\ref{fig:main_hrd} with a purple line.
The fit has the form: \logL\ = 0.47~\Teff\ $-$5.42\,[dex], where \Teff\ is given in kK.

Figure~\ref{fig:vsini_teff} shows that for stars with \logL\,<\,5.0\,dex the adopted position of the drop is consistent for stars within 1500\,pc and 4000\,pc.
Regarding the stars with \logL\,>\,5.0\,dex up to 1500\,pc, we argue that despite the higher degree of completeness, the statistical number of stars is not sufficient to show the drop.
For stars with \logL\,>\,5.35\,dex within 4000\,pc, it can be seen that the drop is somewhat shifted toward cooler temperatures. This could be the result of missing a non-negligible number of stars that are intrinsically fainter on the hot side of the drop, as mid-B-type stars are intrinsically brighter than their early-B and O-type companions \citep[in other words, caused by the Malmquist bias;][]{malmquist22}.

Regarding the distribution of the stars excluded from our sample (see Sect.~\ref{subsection:21.spec_sample}), we expect Be stars to only populate the lower panels of Fig.~\ref{fig:vsini_teff}, since classic Be stars mostly comprise early-B MS objects with luminosity classes III to V, and to do it evenly with \Teff\ \citep[see, e.g.,][]{arcos18}. On the other hand, SB2+ systems represent an important fraction in the O-star domain. For example, \citet{holgado22} showed that $\approx$30\% of their sample of O-type stars are SB2 systems \citep[see also][]{barba17, sana17}, while in \citet[][see Fig.~6]{deburgos23a} we found that the percentage among early- to mid-B-type supergiants drops below $\approx$10\%. The lack of SB2+ systems could for instance explain the apparently lower number of objects above 30\,kK shown in Fig.~\ref{fig:main_hrd} (see also Sect.~\ref{subsection:52_binaries}). 
In addition, the lower number of objects around 32\,kK with \logL\,>\,5.0\,dex (more clearly shown in Fig.~\ref{fig:vsini_teff}) could be explained considering that O9-type stars, which populate that parameter space, gather the largest number of detected SB2 systems. However, a proper diagnostic of this feature would also require a better understanding on how binary interaction products populate the HR diagram.
We also note that the left-skewed distributions shown in Fig.~\ref{fig:vsini_teff}, particularly for stars with \logL\,<\,5.0\,dex, partially result from the effect of stars reaching different maximum \Teff\ values along the ZAMS.


\subsection{Spin-rates of the sample}
\label{subsection:42_spinrates}

\begin{figure}[!t]
 \centering
    \includegraphics[width=0.42\textwidth]{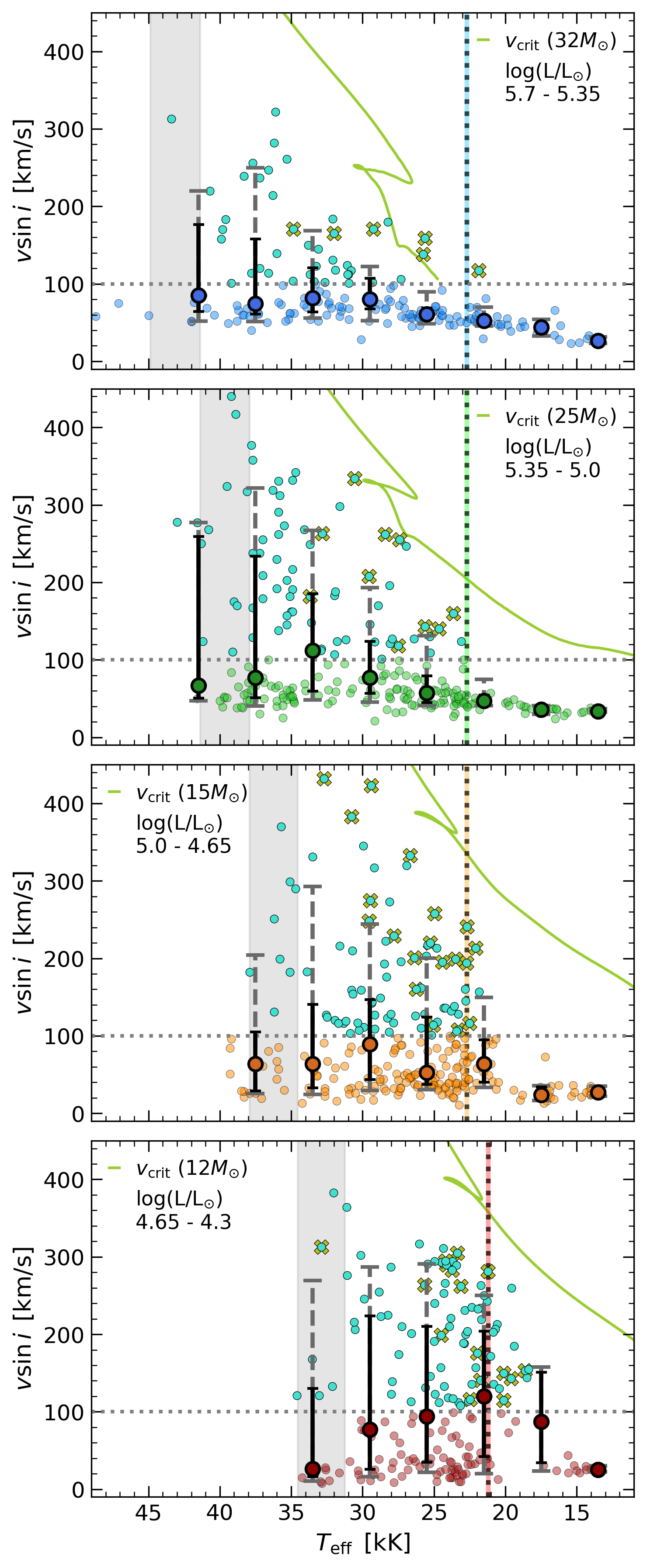}
    \caption{Projected rotational velocities of the stars in our sample against their effective temperature. In this case, the sample is not limited to any distance. The panels separate stars into different luminosity ranges as in Fig.~\ref{fig:vsini_teff}, following the same color code.  
    Solid black error bars and central circles indicate the 75\% and 25\% percentiles and the median values of the stars in bins of 4\,kK. The dashed gray error bars correspond to the 90\% and 10\% percentiles. The dotted line at 100\kms separates fast-rotating stars. The location of the TAMS is also included with a vertical dotted line. 
    The solid green lines correspond to the critical velocity calculated from the models in Fig.~\ref{fig:main_hrd}.
    Stars showing double sub-peak emission in H$\alpha$ are marked with additional lime-colored crosses (see Appendix~\ref{apen.disks_vcrit}). 
    The gray area corresponds to the temperature range covered by the theoretical ZAMS in each luminosity range.}
    \vspace{-0.5cm}
\label{fig:scatter_vsini_teff}
\end{figure}

Another observational characteristic proposed to delineate the location of the TAMS is the drop of the upper envelope of the \vsini\ distribution as a function of \Teff\ \citep[see][]{vink10, brott11}. 
Figure~\ref{fig:scatter_vsini_teff} shows our distribution of \vsini\ values against \Teff\ separating stars in the same luminosity ranges as in Fig.~\ref{fig:vsini_teff}. To qualitatively evaluate the aforementioned feature, we included all the stars regardless of their distance.
The figure indicates the location of the ZAMS (gray areas), the proposed empirical TAMS (dashed colored lines), and outlines the critical rotational velocity (\vcrit; solid green lines), that is, the equatorial velocity at which the centrifugal force balances gravity, as a function of \Teff\ (see below).
We also include the median \vsini\ values for stars within bins of 4\,kK (black-edge colored dots), and the associated 25\% and 75\% (solid black lines) and 10\% and 90\% (dashed gray lines) percentiles. 
Although a detailed discussion of the connection with the TAMS is presented in Sect.~\ref{subsubsection:511_lack_FR}, here we highlight the three most interesting features of Fig.~\ref{fig:scatter_vsini_teff} (see also Appendix~\ref{apen.min-vsini_logL}).

A first feature is the separation of the stars into two main components (see also Fig.~10 of \citet{deBurgos24a}): a low-\vsini\ densely populated group of stars (colored differently) and a less populated group of fast-rotating objects (cyan circles; indicated also in Fig.~\ref{fig:main_hrd}). Following \citet{deBurgos24a}, we adopt the value of \vsini\,=100\kms that best separates both components. This separation is well known to be present in the O-star domain and extends toward the early-B supergiant domain \citep[see, e.g.,][and references therein]{holgado22, deburgos23a}. 
Using as reference those stars within 2500\,pc, we find a decreasing fraction of fast-rotating objects from $\approx$50\% for stars with luminosities below 4.65\,dex, to $\approx$35\% for stars between 5.0 and 4.65\,dex, and to $\approx$30\% for stars with luminosities higher than 5.0\,dex. On average, 38\% of the stars located on the hot side of the TAMS belong to the fast-rotating component. 
A further discussion of both components can be found in Sect.~\ref{subsubsection:512_evol-vsini} and Sect.~\ref{subsection:53_stell-ev}.

Another feature is the absence of fast-rotating objects below a certain \Teff, located very close to the TAMS, which is especially evident in the middle panels. This can also be seen in Fig.~\ref{fig:vsini_teff}, where fast-rotating stars are indicated in cyan bins in all histograms. 
In addition, the \Teff\ that outlines the lack of these objects seems to shift with luminosity by a few thousand kelvins, with fast-rotating stars extending beyond the TAMS (in evolutionary terms) toward lower luminosities and vice versa. As shown in Sect.~\ref{subsection:53_stell-ev}, this behavior is similar to what is predicted in some evolutionary models of stars born with half of their critical velocity.

A third characteristic is the consistent decrease of the maximum \vsini\ reached by stars with decreasing \Teff\ within each luminosity range. In fact, the upper envelopes closely follow \vcrit, which is shown with a solid green line (shown also in the right panel of Fig.~\ref{fig:main_hrd}) and has been calculated from the same models used in Fig.~\ref{fig:main_hrd} employing Eq.~2 in \citet{langer98}.
All these features are discussed in more detail in Sect.~\ref{subsection:51_spinrates}.


\subsection{Comparison with evolutionary tracks}
\label{subsection:43_stell-ev}

\begin{figure}[!t]
 \centering
    \includegraphics[width=0.46\textwidth]{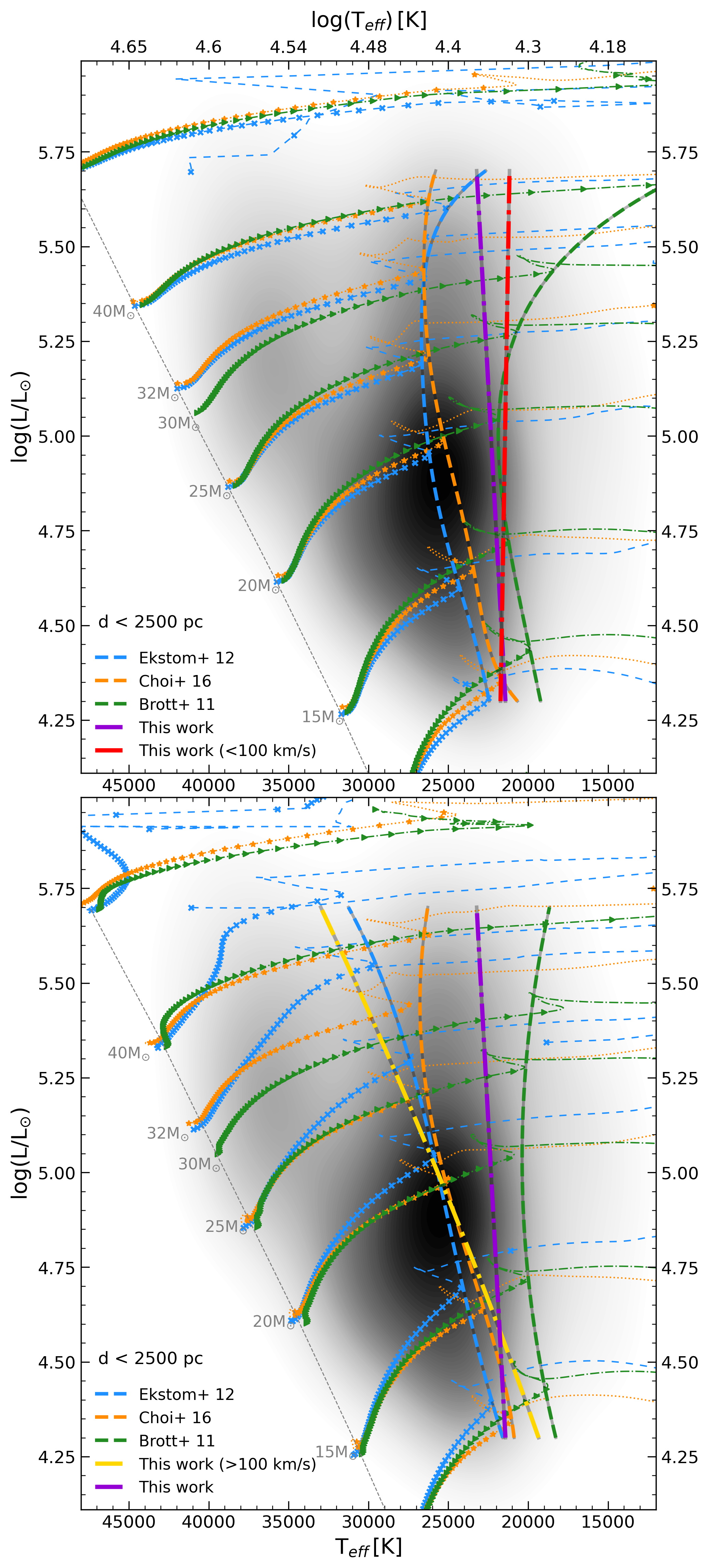}
     \caption{Hertzsprung-Russell diagrams showing different evolutionary tracks for solar metallicity together with different predictions for the location of the TAMS as indicated in each legend. The top panel includes tracks with no initial rotation, and in the bottom panel the tracks correspond to $v_{\rm ini}$/\vcrit\,$\approx$0.4. Intervals of the same age difference are marked with different symbols. The gray density mesh in the background is based on a Gaussian kernel and includes our sample stars up to 2500\,pc. Evolutionary tracks are limited up to the RSg phase (if applicable).}
\label{fig:models}
\end{figure}

Three publicly available evolutionary calculations widely used in the literature are those of \citet{brott11}, \citet{ekstrom12}, and \citet{choi16b}. Each uses different assumptions, prescriptions, or approximations to control internal physical processes including, among others, angular momentum transport from the core to the surface, rotational mixing, or core overshooting. Figure~\ref{fig:models} displays the evolutionary tracks of these models without initial rotation (top panel) and with considerable initial rotation\footnote{Models from \citet{brott11} were chosen to have $v_{\rm ini}$/\vcrit\,$\approx\,$0.4 among the different available $v_{\rm ini}$ values.} ($v_{\rm ini}$/\vcrit\,=0.4; bottom panel). 
To compare our proposed empirical TAMS (dashed-dotted purple line) with the equivalent TAMS of evolutionary models, we performed a polynomial fitting to those \Teff\,-\,\logL\ points on the coolest side of the hook toward the end of the MS \citep[see Sect.~\ref{section:1_tmp} and][]{martinet21}.
Since our empirical TAMS mixes stars with different rotations, the top panel includes for reference an additional TAMS where we removed all fast-rotating stars (dashed-dotted red line; hereafter SR-TAMS). This results in an almost vertical line at \Teff\,=\,22.65\,kK ($\sigma$\,=\,0.11\,kK).
Complementary, the bottom panel includes an alternative empirical TAMS when we limit out sample to only fast-rotating objects (dashed-dotted yellow line; hereafter FR-TAMS). In this case, the second-order polynomial fit follows the form:
\logL\ = $-$0.001~\Teff$^{2}$ + 0.154~\Teff\ +1.71\,[dex], where again \Teff\ is in kK.

Starting with the top panel, the comparison with the models by \citet{ekstrom12} and \citet{choi16b} shows that in the 15\,--\,30\MSol range, both predict a hotter effective temperature for the TAMS. However, both agree well below the 15\MSol. 
In contrast, the TAMS predicted by \citet{brott11} is closer to ours up to the 25\MSol track, but significantly extends the width of the MS above it. We note that this extended MS, while also present in the other evolutionary calculations, only becomes noticeable above 40\MSol. Unfortunately, the number of stars in our sample is not sufficient to draw any decent comparison.
We also note that for models computed without initial rotation, the difference between our TAMS and SR-TAMS is not substantial (see below).

In the bottom panel of Fig.~\ref{fig:models}, our TAMS only coincides with those of \citet{ekstrom12} and \citet{choi16b} for masses \ls15\MSol. Interestingly, our FR-TAMS overlaps well with both sets of models up to $\approx$25\MSol, and for models by \citet{ekstrom12}, the agreement continues up to 40\MSol. However, models by \citet{ekstrom12} for this initial rotation also predict an important spin-down of the stars near the MS. Therefore, an adequate comparison requires further considerations regarding the evolution of the spin-rate properties of the stars.
The models of \citet{brott11} extend the width of the MS toward cooler temperatures with respect to our TAMS and FR-TAMS in the entire range of covered masses.

We remark here that our definition of the TAMS has been adopted as the temperature that includes 15\% of the coolest observed objects, but this depends, for example, on the number of objects located on the hot side of the TAMS. Therefore, the above comparison is affected by this definition. Despite this, the general comparison with the different models indicates that there is no definitive set of models that can reproduce our empirical TAMS in the full range of masses (see Sect.~\ref{subsection:53_stell-ev}, where we discuss the reasons for the observed discrepancies).


\subsection{Distribution of SB1 objects in the HR diagram}
\label{subsection:44_SB1}

\begin{figure}[!t]
 \centering
    \includegraphics[width=0.46\textwidth]{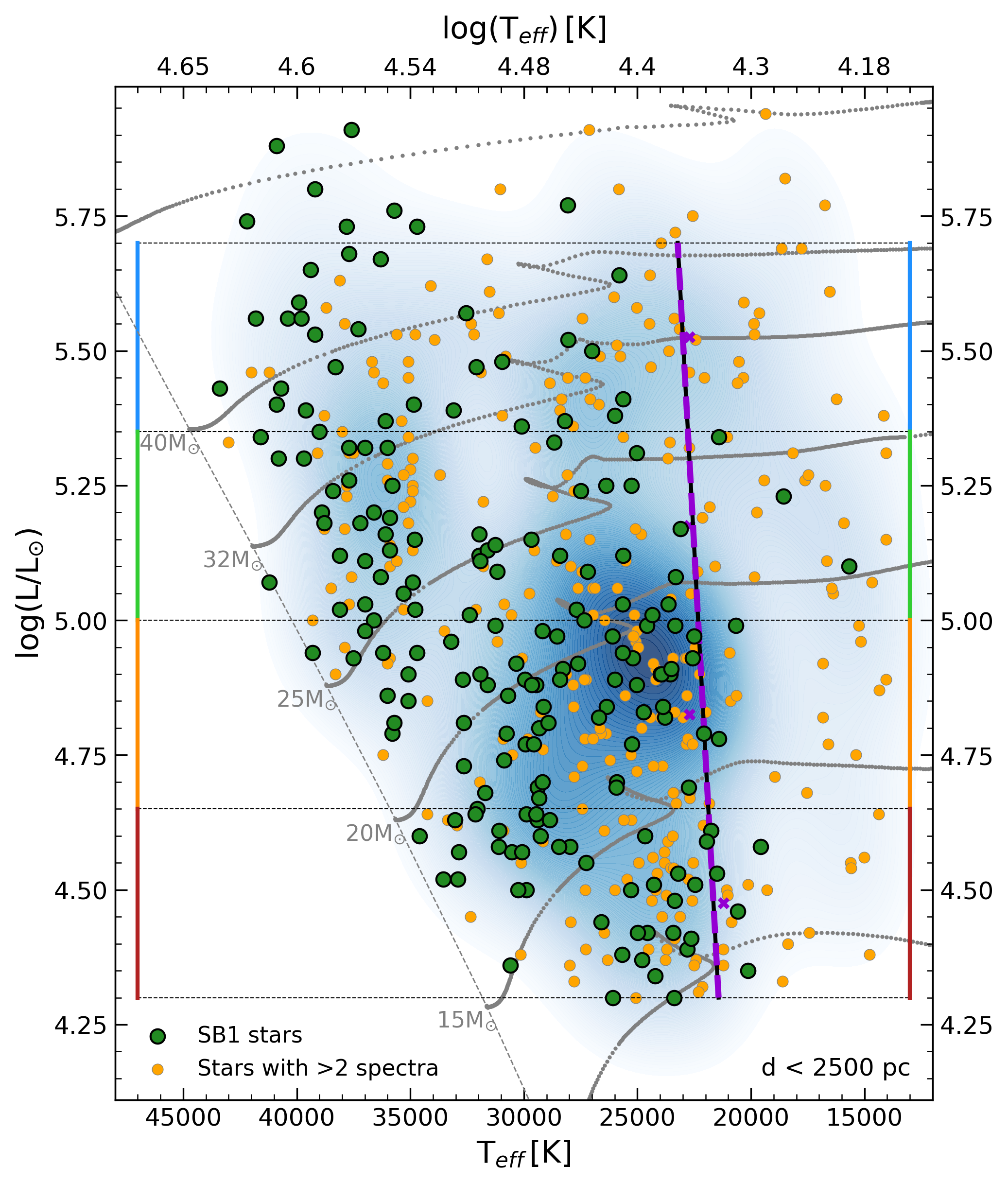}
    \caption{Similar HR diagram as in Fig.~\ref{fig:main_hrd} but limiting the sample to stars within 2500\,pc with three or more available multi-epoch spectra. Green circles highlight those stars identified as SB1.}
\label{fig:sb1_hr}
\end{figure}

As explored in \citet{mcevoy15} using a sample of BSGs in the 30~Doradus region, the lack of SB1 systems in the HR diagram can also provide additional constraints on the location of the TAMS. Moreover, the lack of binaries on the cool side of the TAMS can provide clues about the nature of those objects located beyond the MS.

We followed the guidelines presented in \citeauthor{simon-diaz24a} \citep[\citeyear{simon-diaz24a}; see also][]{holgado18} to separate the observed peak-to-peak amplitudes of the radial velocity caused by the intrinsic variability of the stars from true SB1 systems. In particular, the work by \citet{simon-diaz24a} establishes \dvrad\ thresholds with the effective temperature and the spectroscopic luminosity\footnote{The spectroscopic luminosity \Ls, is defined as $T_{\rm eff}^4$\,/$g$ \citep[see][]{langer14}} for this separation (shown in Appendix~\ref{apen.rvpp_thresholds}), in addition to several visual guidelines and a minimum number of multi-epoch data to avoid erroneous detections.

We evaluated the presence of SB1 systems in our sample limited to a distance of 2500\,pc. Figure~\ref{fig:sb1_hr} shows the location in the HR diagram of the 494 systems with three or more available spectra, where we identified 176 SB1 systems (green circles). It can be seen that they are spread mainly within the MS band of the HR diagram, with only a few stars beyond the position of the TAMS. 
A further discussion of this difference is included in Sect.~\ref{subsection:52_binaries}.


\subsection{Comparison with Castro et al. (2014)}
\label{subsection:45_castro}

This result represents an important improvement over the previous reference work by \citet{castro14}, who used a compilation of spectroscopic observations and results from the literature to show the distribution of 439 massive OB stars, for the first time, in a spectroscopic HR (sHR) diagram, substituting the luminosity of the traditional HR diagram by the spectroscopic luminosity. 
Comparing our results with those of \citet{castro14}, we first point out that despite the fact that part of their sample populates the same parameter space as ours \citep[see][]{deBurgos24a}, their sample of stars was not created homogeneously and included targeted groups of stars of interest for specific studies. Our sample also benefits from being analyzed in a homogeneous self-consistent manner, rather than being composed of heterogeneous results available in the literature.

Additionally, observational biases due to the different intrinsic brightnesses of the stars are likely present in their work, whereas our volume-limited sample with high degree of completeness solves this issue. 
Furthermore, while the sHR diagram is used in \citet{castro14}, here we make use of the HR diagram, combined with additional information on the spin-rate properties and binary fraction of the stars (Sects.~\ref{subsection:42_spinrates} and \ref{subsection:44_SB1}).
Moreover, we consider double the number of stars in \citet{castro14} up to 2500\,pc within the same \Teff\ range. We also note that in \citet{castro14}, the surface gravities were not corrected by centrifugal forces \citep[see][]{herrero92, repolust04}, shifting fast-rotating stars toward apparently higher luminosities in the sHR diagram. 

Despite the differences, \citet{castro14} found a drop in the density of stars below 25\MSol at \Teff\,$\approx\,$20\,kK, which was also interpreted as the TAMS. However, for stars with 25\,--\,40\MSol\ the drop was located at \Teff\,$\approx\,$10\,kK.
As shown in Figs.~\ref{fig:main_hrd} and \ref{fig:vsini_teff}, our results indicate that the drop occurs throughout the full range of luminosities and masses considered here, and it is located approximately at the same \Teff\ (24\,--\,22\,kK). 
The overdensity of stars shown in \citet{castro14} beyond 20\,kK, which is not observed in any of the panels of Fig.~\ref{fig:main_hrd}, is likely the result of a selection bias introduced by considering any star with available parameters from the literature rather than a volume-limited sample. In particular, since mid-B-type stars are intrinsically brighter in the optical compared to early-B- and O-type objects, the aforementioned overdensity is likely caused by mid-B-type stars located at larger distances.


\section{Discussion}
\label{section:5_tmp}


\subsection{Spin-rate properties of the sample}
\label{subsection:51_spinrates}

Rotation in massive stars plays an important role in their entire evolution \citep{meynet00, heger00, maeder05, yoon05, langer12}. 
It can enhance internal mixing processes that provide the stellar core with additional hydrogen, allowing stars to extend their time in the MS. Rotation can also trigger increased mass-loss rates, which can affect their evolution and final fates.
Therefore, understanding the observed spin-rate properties of massive stars is necessary to better constrain their evolution.

As indicated in Sect.~\ref{subsection:42_spinrates}, two main observable features are known to characterize the spin rates of luminous O- and B-type stars. One is the bimodal distribution of the projected rotational velocity \citep[e.g.,][]{conti77, howarth97, dufton13, ramirez-agudelo13, holgado22, deBurgos24a}, which includes a main component of slowly rotating stars and a secondary more extended tail of fast-rotating objects. The latter have been proposed to be the products of binary interactions \citep[e.g.,][]{ramirez-agudelo13, demink14, ramirez-agudelo15, holgado22, britavskiy23, deBurgos24a}.
The second feature is the lack of fast-rotating stars below a certain temperature \citep[see][]{howarth97, ryans02, vink10, simon-diaz14a, mcevoy15, deBurgos24a}. 
As indicated by different studies, this characteristic could be interpreted as the empirical location of the TAMS \citep[see][]{vink10, brott11, martinet21, deBurgos24a}. We explore this possibility further.


\subsubsection{The lack of fast-rotating stars beyond the TAMS}
\label{subsubsection:511_lack_FR}

Previous representations of the \vsini\ distribution against \Teff\ have shown the presence of a clear drop from 300\,--\,200\kms to 70\,--\,30\kms of the most rapidly rotating objects around 22\,--\,20\,kK \citep[see][]{vink10, mcevoy15, deBurgos24a}.
The distributions shown in Fig.~\ref{fig:scatter_vsini_teff} for stars with \logL\,<\,5.35\,dex also evidence this feature.

On the one hand, massive stars leaving the MS should undergo an expansion of their envelopes and, in turn, also decrease their observed surface rotational velocities. However, this expansion is not sufficient to drastically reduce spin rates within the investigated \Teff\ range where the drop occurs (which is of $\approx$200\kms; see, e.g., the second panel of Fig.~\ref{fig:scatter_vsini_teff}). 

On the other hand, stars with expanding envelopes and high spin rates near the TAMS can get too close to their critical rotational velocities \citep[see, e.g.,][]{meynet06, meynet07}. It is predicted that before reaching that point, stars expel enough material into the medium \citep[e.g.,][]{kurfurst14}, triggering a spin-down of the star's rotation \citep[e.g.,][]{georgy11, demink13}.
As shown in Fig.~\ref{fig:scatter_vsini_teff} (see also the right panel of Fig.~\ref{fig:main_hrd}), the critical rotation that we extract from the models approximates the upper envelope of the \vsini\ distribution, particularly in the two central panels. We note that, except for the top panel, we could, in principle, expect to find some fast-rotating objects that have not yet reached their \vcrit. 

A third possibility is related to mass loss. In this respect, \citet{vink10} suggested that the bi-stability nature of the winds \citep[see][]{pauldrach90, vink99, vink00} can lead to an important increase in the mass-loss rates at $\approx$22\,kK, and to a subsequent enhancement of the angular momentum loss (known as the bi-stability braking). However, in \citet{deBurgos24b} \citep[see also][]{rubio-diez22, krticka24, bernini-peron24} it is shown that such an increase in mass-loss rates (by a factor of ten to twenty) is not observed\footnote{Newer prescriptions, such as those of \citet{bjorklund21}, do not predict any enhancement near our empirical TAMS.}. Therefore, we argue that the lack of fast-rotating stars is not due to the latter proposed mechanism.

Given these considerations, the lack of fast-rotating objects could be the result of the statistically and relatively lower number of stars expected beyond the TAMS, that is, the chances of observing these objects is lower. This could be further supported by the following. Considering that for stars with \logL\,=\,5.35\,--\,4.65 (where we account for more stars), the histograms of Fig.~\ref{fig:vsini_teff} show the largest difference in the relative number of stars when crossing the TAMS, which we locate at an average of 22.5\,kK. For stars up to 2500\,pc, there are five times fewer stars between those located on the hot and the cool sides of the TAMS within 5\,kK.
Considering that the proportion of fast-rotating objects located within each side with respect to the total is 1:3, this exercise would result in only six fast-rotating objects below 22.5\,kK. We account for two objects that meet these conditions, which we consider sufficient to support the aforementioned possibility. We note that we do not account for the fact that some of the fast-rotating stars may not be seen as such because of a projection effect.
Furthermore, if fast-rotating stars are the product of binary interaction, it is expected to occur mainly when at least one of the components of the binary system is still on the MS \citep[see][]{demink13, demink14}.
Therefore, it would be difficult to observe these objects beyond the TAMS. Additionally, fast-rotating objects evolving near or beyond the TAMS will eventually reach their \vcrit, thus becoming slowly rotating stars.
Nonetheless, one cannot rule out the possibility of unconsidered effects or forces that can trigger moderate braking toward the end of the MS. This can include, for example, the effect of magnetic fields \citep[see][]{keszthelyi20, keszthelyi22}.

In conclusion, we argue that the lack of fast-rotating stars below a given temperature is unlikely to be caused by enhanced mass-loss rates and may have its origin in two possible compatible explanations. One, that beyond the TAMS, the statistical number is not sufficient to observe them; and two, that if fast-rotating objects result from binary interaction within the MS, then it would be difficult to observe them beyond the TAMS. Either way, we argue that the drop of fast-rotating objects can also be used to outline the end of the MS.


\subsubsection{\texorpdfstring{The evolution of \vsini\ along the MS}{The evolution of the vsini along the MS}}
\label{subsubsection:512_evol-vsini}

Figure~\ref{fig:scatter_vsini_teff} provides the largest \vsini\ distribution of Galactic OB stars to date, which is combined with the largest \Teff\ and $L$ coverage of stars homogeneously analyzed. This allows us to confidently extend our discussion of how the stars within each of the two \vsini\ components may evolve along the MS. 

Interestingly, the median \vsini\ values of the two top panels in Fig.~\ref{fig:scatter_vsini_teff} show similar trends toward lower effective temperatures. The values are approximately constant until 30\,kK and then slowly decrease toward the cool end. This trend is less obvious in the third panel, but may still be present, and it is certainly not the case of the bottom panel, which we discuss separately in Appendix~\ref{apen.vsini_4.7-4.3}.
We also note that the first median value may not be so representative, as it overlaps with the position of the ZAMS, where the number of stars is significantly lower than for the other mean values. 
Excluding that one, our results strongly suggest a slow-braking process occurring along the MS, which has already been empirically supported by the works of \citet[][]{holgado22, deBurgos24a}.
Furthermore, the number of stars on the hot side of the TAMS with \vsini\,\ls100\kms (i.e., the low-\vsini\, component of the spin-rate distribution) represents 65\% of the total number. This large fraction may indicate that massive stars are born with low or mild initial rotations (not more than 0.2 of the critical velocity), as indicated also in the aforementioned works. However, we note that since we do not account for projection effects, that percentage may represent a lower limit.

In Sect.~\ref{subsection:42_spinrates} we showed the moderate decrease of the highest \vsini\ values with decreasing \Teff\ until the TAMS is reached (see Fig.~\ref{fig:scatter_vsini_teff}; see also Sect.~\ref{subsubsection:511_lack_FR}). In general, these values follow the limit imposed by the critical velocity, indicating that fast-rotating objects are ubiquitous within the MS. 
In this regard, the 90\%-percentiles associated with the median values shown in Fig.~\ref{fig:scatter_vsini_teff} also provide a better statistical representation of this fact.
We also note here that the percentile corresponding to the first median value is not representative, whereas the rest of the values monotonically decrease toward lower temperatures.
We noticed that the 90\%-percentiles associated with the top panel (stars with \logL\,=\,5.7\,--\,5.35\,dex) are more separated from the \vcrit\ than in the other panels. A possible explanation for this could be related to the importance of mass loss with increasing luminosity and its connection with angular momentum loss \citep{maeder09}. In this regard, \citet{deBurgos24b} showed that the difference in mass-loss rates between BSGs with \logL\,<\,5.0\,dex and those with \logL\,>\,5.3\,dex can be more than a factor of ten.
If this is the case, stars with higher luminosities would lose angular momentum at a higher rates than their less luminous counterparts.


\subsection{The impact of binaries}
\label{subsection:52_binaries}

An important aspect to our understanding of massive stars is the impact that binary interaction has on the interpretation of the observed properties. This is particularly relevant as a result of the large proportion of objects that are predicted to interact with a companion during the MS \citep[see][]{sana12, demink14}.

Regarding the detection of SB1 binary systems, it is important to remark that there is no definitive method to distinguish \vrad\ variations produced by stellar companions from those caused by intrinsic stellar variability \citep[see][for a review]{bowman20a}. Moreover, this variability can have different physical origins, including pulsations that originate in the stellar interiors or strong stellar winds that originate in their atmospheres \citep[see, e.g.,][]{simon-diaz10b, aerts17, aerts18, burssens20, simon-diaz24a}.
Therefore, the percentage of detected SB1 systems is highly dependent on the adopted threshold used to separate the effect of intrinsic variability from the variations linked to orbital motion.
In this work, we followed the approach of \citet{simon-diaz24a}, which provides a study of the characteristic amplitude of \vrad\ variations produced by stellar pulsations in the O-stars and BSGs domains, allowing us to more confidently indicate the likely presence of a stellar companion.

In Sect.~\ref{subsection:44_SB1} we present the distribution of detected SB1 systems in an HR diagram, limited to a distance of 2500\,pc. Figure~\ref{fig:sb1_hr} shows the proposed TAMS and the lack of SB1 systems beyond it, on the cool side. By evaluating the fraction of SB1 systems on each side of the TAMS, we obtain 39\% within \Teff\,=\,30\,--\,22.5\,kK and 15\% within \Teff\,=\,22.5\,--\,15\,kK, highlighting an important decrease when crossing the TAMS. Interestingly, the average percentage is similar to the 23$\pm$6\% found by \citet{dunstall15} in the 30 Doradus region of the Large Magellanic Cloud. Furthermore, our results approach those found by \citet{mcevoy15} in the Tarantula Nebula, where SB1 systems were not detected below a given temperature. In this work, we show that SB1 systems are also present beyond the TAMS, but their number is very low. We conclude that the lower percentage of SB1 systems beyond the TAMS can be used to infer its location with respect to \Teff. 
We also argue that the few SB1 systems beyond the TAMS likely correspond to stars heading to the RSg phase, since this phase practically eliminates all detectable binaries beyond the TAMS \citep[see][]{demink14}.

One of the predicted types of binary interaction involves that one of the stars in the system reaches a more evolved state in which its larger size leads to a mass transfer into the companion. The donor can then become a compact object (i.e., a black hole or neutron star) or a stripped helium star. This interaction may result in an SB1 system, a merge, or in the primary star being kicked by supernova explosion of the secondary. The latter two cases do not produce an SB1 system.
In either case, this type of interaction is less likely to occur outside the MS, as it would require both stars to have similar masses to reach the end of the MS at the same time \citep[see][]{marchant24}. 
Therefore, one would expect the number of SB1 systems to decrease significantly beyond MS, not only in relative terms (caused by the fast post-MS evolution), but also as a result of this less probable scenario. This situation is supported by our findings. We refer to Simón-Díaz et al. (in prep.) for a detailed evaluation of the evolution of the fraction of SB1 and SB2+ systems across the MS.

Our results also indicate that the percentage of SB1 systems within \Teff\,=\,31\,--\,22.5\,kK decreases in the upper half of the HR diagram. In particular, from 40\% for stars with \logL\,<\,5.0 to 31\% for stars above that luminosity. This case could be partially explained by the higher detection threshold of \vrad\ (up to 25\kms; see Appendix~\ref{apen.rvpp_thresholds}) toward our highest luminosities, which would eliminate some true SB1 systems.

\begin{figure}[!t]
 \centering
    \includegraphics[width=0.47\textwidth]{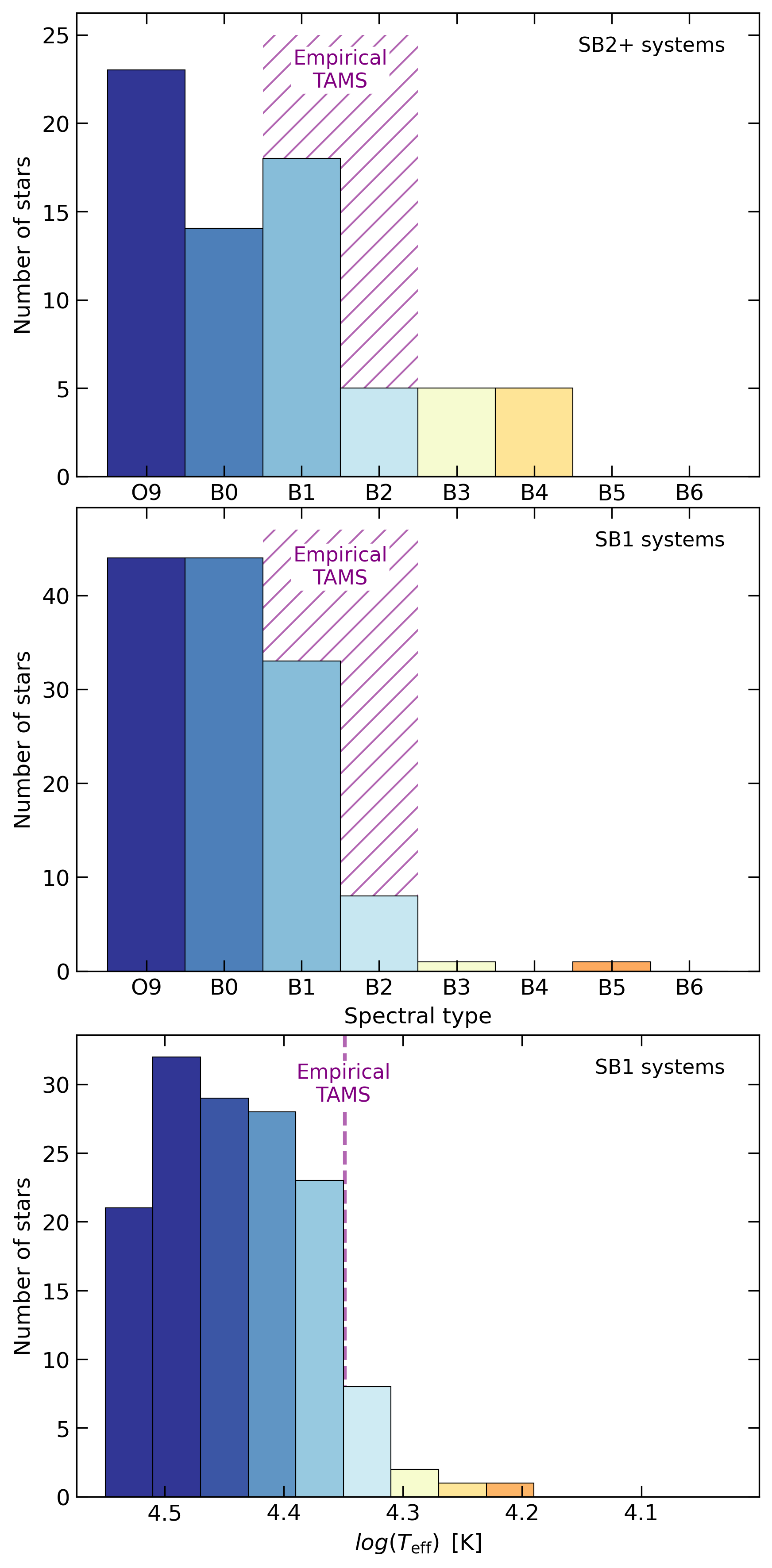}
    \caption{Histograms of the number of spectroscopic binaries against spectral type or $log_{10}(T_{\rm eff})$, limited to a distance of 2500\,pc. The top panel include SB2+ systems for which the spectral type is taken from the primary star (see also Sect.~\ref{subsection:52_binaries} for further details). The middle panel is similar to the top one but including the SB1 systems detected from the \dvrad\ variability. The bottom panel include the same stars as in the middle one, but we used the $log(T_{\rm eff})$ instead of the spectral type. In the top two panels, the position of the TAMS is highlighted with a patch of purple lines, derived from the range of spectral types of stars within 1\,kK of the average \Teff\ of the TAMS, which is shown in the bottom panel as a dashed purple line.}
\label{fig.SB1-2}
\end{figure}

Similarly as for SB1 systems, the presence of a drop in SB2+ systems on the cool side of the TAMS could provide additional evidence of its location. Although SB2+ systems were intentionally excluded from our sample, and therefore, we do not count with reliable quantitative spectroscopic analysis of each individual component, we can still use to some extent the available information on their spectral types as a proxy of the corresponding effective temperatures.

The upper and middle histograms of Fig.~\ref{fig.SB1-2} show the distribution of SB2+ and SB1 systems with respect to their spectral type\footnote{The spectral classification was adopted from trustable sources when possible, or from the SIMBAD astronomical database.}. In this case, SB2+ systems are also limited to a distance of 2500\,pc. SB2+ systems are limited to B-type supergiants (a primary component of luminosity classes I and II) and O9-type stars (of any luminosity class), or with a \fwhb\,<\,7\AA. As shown in \citet{deburgos23a}, the quantity \fwhb\ is defined as the difference between the width of the H$\beta$ line measured at three-quarters and one-quarter of the depth of the line and can be used as a proxy of \Ls\ to select BSGs independently of their spectral classification and \vsini. The selected SB2+ systems include those identified in \citet{deburgos23a}, plus some others found later from the acquisition of new spectra. The SB1 systems are limited to stars in our sample with spectral types O9\,--\,B6.

Interestingly, SB1 and SB2+ systems follow a similar overall distribution with respect to their spectral types: a high number of systems from O9 to B1 spectral types, followed by a steep drop at the B2 spectral type. In both cases, this drop roughly coincides with the empirical location of the TAMS (patched areas with diagonal purple lines). 
To provide a rough idea of the comparison between \Teff\ and spectral types, the bottom histogram of Fig.~\ref{fig.SB1-2} shows this equivalency, where the drop of SB1 systems very well matches the position of the TAMS at $log_{10}(T_{\rm eff})$\,=\,4.35\,dex ($\approx$22.5\,kK; dashed purple line).

These results also provide evidence that the number of SB2+ systems in the BSG domain before reaching the TAMS is lower than for O9 stars. In this regard, given their relatively low number, the lack of SB2+ systems in the sample should not produce important biases with respect to the distribution of stars in Figs.~\ref{fig:main_hrd}, \ref{fig:vsini_teff}. However, the difference between O9- and B0-type stars could partially explain the apparent lower density of objects around 32\,kK (see Sect.~\ref{subsection:41_drop}). We also notice the larger number of B3\,--\,B4 SB2+ systems compared to SB1 and we refer to Simón-Díaz et al. (in prep.) for further discussion.

From this exercise, we conclude that the drop in SB2+ systems coincides with that of the SB1 systems and could also be used as an indicator of stars reaching the TAMS. In fact, we also expect the number of SB2+ systems to decrease beyond the MS as a result of (at least) one of the stars in the system either turning into a compact object or interacting with the companion.


\subsection{Implications for stellar evolution models}
\label{subsection:53_stell-ev}

The drop in density in the HR diagram and the lack of fast-rotating stars and SB1 systems beyond that drop all support the location of a new empirical TAMS at the position described in Sect.~\ref{subsection:41_drop}.
In Sect.~\ref{subsection:43_stell-ev} we show that this position implies an extension or contraction of the width of the MS depending on the set of evolutionary models considered.
Moreover, we showed that the location of the TAMS depends on the selection of stars based on their \vsini. In this section, we further discuss the comparison with widely used publicly available stellar evolutionary models. 

We argue in Sect.~\ref{subsubsection:512_evol-vsini} that one possibility to explain the observed \vsini\ distribution and the larger fraction of stars with \vsini\,\ls100\kms is that massive stars are born with low to mild initial rotation ($v_{\rm ini}$\,\ls150\kms) and continue evolving while slowly decreasing their spin rates. This scenario is more compatible with the predictions of \citet{brott11} and \citet{choi16b}, where the effect of surface braking is much lower than in \citet{ekstrom12}. 
Although the models of \citet{ekstrom12} and \citet{choi16b} do not include an intermediate initial rotation ($v_{\rm ini}$/\vcrit\,\ls0.2) we continue our discussion assuming that the available models without initial rotation (top panel of Fig.~\ref{fig:models}) are more representative to compare the positions of the theoretical and empirical TAMS.
In this regard, two important processes that significantly impact the width of the MS across all the masses considered here are the above-mentioned initial rotation and the amount of overshooting.
Other mechanisms such as semi-convection, inflation, mass-loss rates, rotationally induced chemical mixing, and magnetic fields also influence the MS, some of which are restricted to a specific range of masses \citep[e.g., inflation for masses \gs30\MSol; see][]{ishii99, kohler15}, or their influence also depends on additional assumptions \citep[e.g., rotational mixing which relies on an efficient angular momentum transport; see][]{martinet21}. As shown by \citet{keszthelyi22}, the effect of surface fossil magnetic fields can largely influence the evolution of stars and certainly deserves further consideration. However, the models of \citet{brott11}, \citet{ekstrom12}, and \citet{choi16b} do not account for this effect; therefore, we decided not to include it in our discussion. 

This leaves the main difference for the position of the TAMS in Fig.~\ref{fig:models} set by the overshooting parameter ($\alpha_{ov}$). In the models of \citet{brott11}, this parameter is fixed to $\alpha_{ov}$\,=0.335, whereas in \citet{ekstrom12}, it takes the value of $\alpha_{ov}$\,=0.10. For \citet{choi16b}, they used the exponential overshoot formalism, which for models without initial rotation is equivalent to $\alpha_{ov}$\,$\approx\,$0.20\footnote{We note that this comparison is not straightforward in models with rotation because it induces diffusive processes that contribute to chemical mixing and thus the length of the MS.}.
In general, in all available models, the coefficients that regulate this parameter have been set to match either the empirical location of the drop in the relative number of stars \citep[e.g.,][]{ekstrom12}, or the drop in rotational velocities \citep[e.g.,][]{brott11}. 
The latter would be more risky under the aforementioned assumption of a low initial rotation, plus the additional uncertainty in the case that fast-rotating objects have their origin in binary interaction (see Sect.~\ref{subsection:52_binaries}). 
Additionally, the range of available masses used to calibrate the coefficients in those models where overshooting is not mass-dependent is an additional source of discrepancy. 
This explains, for example, the better agreement between the models by \citet{brott11} and \citet{ekstrom12} for intermediate masses below $\approx$15\MSol.

Recent work by \citet[][]{martinet21} used the observational information presented in \citet{castro14} to investigate the sizes of the convective cores and constrain the amount of overshooting. 
This and other works have also shown that the size of the cores is mass-dependent \citep[e.g.,][]{langer14, claret19, tkachenko20, tkachenko24}. 
Using our volume-limited sample and its high degree of completeness, we can contribute to a better definition of the width of the MS by adding new constraints to fine-tune the overshooting parameter and its dependence on the initial mass for evolutionary masses from 12\MSol and up to 40\MSol.
In this regard, we show in Sect.~\ref{subsection:43_stell-ev} that our empirical TAMS lies between the theoretical TAMS of \citet{ekstrom12} and \citet{choi16b}, and the one of \citet{brott11}. Although more in-depth comparisons are certainly required, our results suggest higher overshooting values in the first two cases and a lower value above 25\MSol in the latter case.
Moreover, the lack of an overdensity of objects with 25\,--\,40\MSol below $\approx$20\,kK as shown in \citet{castro14} (see Sect.~\ref{subsection:45_castro}) also adds an important constraint for future models. Although some authors attributed the observed extension to the effect of inflation \citep[e.g.,][]{choi16b, sanyal17}, this is no longer needed within that range of masses.
Unfortunately, the number of stars in our sample is not sufficient to draw conclusions above 40\MSol.

In Sect.~\ref{section:1_tmp}  we mentioned the possibility that stars undergoing a blueward evolution may also be present in our sample. If this is the case and some of these stars overlap with those used here to define the position of the TAMS, it could lead to its incorrect definition. We evaluated this possibility using the models by \citet{ekstrom12} and \citet{choi16b}\footnote{Models by \citet{brott11} are not computed beyond the main-sequence phase and therefore are not considered.}. 
The models of \citet{choi16b} only predict stars evolving blueward for $v_{\rm ini}$/\vcrit\,=0.4 and initial masses below 12\MSol\ and above 40\MSol, whereas in the models of \citet{ekstrom12}, we find their presence for masses below 12\MSol (which corresponds to stars undergoing blue loops) and above 32\MSol or 20\MSol for models without rotation or with $v_{\rm ini}$/\vcrit\,=0.4, respectively.
Assuming the slow initial rotation scheme, the models of \citet{ekstrom12} for masses \gs32\MSol predict stars to quickly move blueward and then spend a non-negligible time ($\approx$0.2\,--\,0.3\,Myr) in the 30\,--\,20\,kK \Teff\ range before reaching their final fates.
If this is the case, then some of the stars in this \Teff\ range could contaminate the distributions shown in Fig.~\ref{fig:vsini_teff} for stars with \logL\,\gs5.3\,dex. 
To further evaluate the nature of these objects (whether they are evolving toward the RSg phase or coming from it), additional clues from the surface chemical composition and the pulsational properties are required \citep[see][]{bowman19b}.

As also mentioned in Sect.~\ref{section:1_tmp}, the binary channel can also contribute to the observed distribution of single stars and SB1 systems used to delineate the TAMS \citep[see, e.g.,][]{menon24}. Unfortunately, our understanding of how different types of binary interaction spread over the HR diagram is still in an early stage. Nevertheless, we may expect that for those cases where the merging of two stars leads to a reconfiguration of the outcome core, the evolution follows the same path as for a more massive single stars, thus not altering the drop in relative number.

The initial stellar rotation also influences the extension of the MS predicted by evolutionary models \citep[see, e.g.,][]{meynet00, brott11, martinet21}. 
Although, as pointed out in Sect.~\ref{subsection:43_stell-ev}, we cannot directly compare our empirical FR-TAMS with the theoretical ones, our large sample compared to previous studies \citep[e.g.,][]{vink10, castro14, mcevoy15} allows us to withdraw one conclusion.
If fast-rotating stars retain their spin rates (regardless of whether they got it as a result of binary interaction, or they were born as such), then rotationally induced effects seem to shorten the width of the MS for higher masses and extend it for lower ones. This behavior could be similar to what predicted by \citet{ekstrom12} for $v_{\rm ini}$/\vcrit\,=\,0.4. However, in that case, the models by \citet{ekstrom12} also predict an important spin-down on the initial part of the MS, which is not consistent with our observations.

The comparison with the different models therefore suggests that a revision of the overshooting is required in all the cases to match our empirical TAMS, which is a conclusion also supported by asteroseismic studies of massive stars \citep[see][]{burssens23}. In the case of \citet{ekstrom12} and \citet{choi16b}, the overshooting should be adjusted to further extend the width of the MS to cooler temperatures for masses between 15 to 30\MSol, whereas models by \citet{brott11} appear to require further adjustments.


\section{Conclusions and future work}
\label{section:6_tmp}

The question of whether BSGs are MS or post-MS objects arose over three decades ago as a result of the observed overdensity of BSGs situated beyond the theoretical location of the TAMS.
To date, we know that a significant fraction of BSGs are likely hydrogen-burning stars and that a significant fraction of their observed population may be binary products. However, we do not have a clear definition of the location of the end of the MS, a situation that gets worse considering that, apart from single MS stars, additional evolutionary channels can also populate the BSGs domain, making a very difficult task to disentangle the different groups.
A key to improving this situation requires high-quality statistically significant volume-limited samples of BSGs analyzed homogeneously.

In this work, we considered 670 spectroscopically observed Galactic O-stars and BSGs. Using the ALS~III catalog as reference, we found that our sample comprises $\approx$60\% of all massive stars within 2500\,pc and $B_{\rm mag}$\,<\,11 that are born as O-type stars and reach the spectral type B6. This results in the most complete sample of Galactic massive stars spectroscopically analyzed to date.
Combining the results of the quantitative spectroscopic analysis with \textit{Gaia} distances, we were able to derive their absolute luminosities. 
This allowed us to represent the sample in the HR diagram and to analyze its properties in the context of stellar evolution.

Our results evidenced a drop in the density of stars below 25\,--\,22\,kK and for luminosities between 5.7 and 4.3\,dex (corresponding to initial masses between 12 and 40\MSol). We interpret this drop as the location of the TAMS. Moreover, this drop is still present when the volume of our sample is reduced to 1500\,pc or increased to 4000\,pc. This is an important improvement over previous studies that attempted to empirically locate the TAMS. We particularly improve the situation for initial masses above $\approx$25\MSol, where the location of the TAMS was suggested to be at \Teff\,$\approx\,$10\,kK based on the lower number of objects in the sHR diagram.

We studied the projected equatorial velocities of the stars in our sample with respect to \Teff\ and \logL. We found an important lack of fast-rotating stars (\vsini\,>\,100\kms) on the cool side of the TAMS, almost outlining its position. The lack of these objects has been suggested to be the result of stars leaving the MS, or due to an important braking effect triggered by an increase in mass-loss rates at $\approx$22\,kK. However, the latest findings on mass-loss rates in BSGs do not support the latter scenario.
We concluded that the lack of fast-rotating objects can be a consequence of the lower relative number of stars expected beyond the TAMS due to rapid post-MS evolution, combined with the fact that those stars possibly reach their critical velocity near the TAMS, triggering a spin-down of their rotation. Furthermore, this scenario is compatible with the possibility that fast-rotating objects are products of binary interaction occurring during their MS phase.
We also find that the \Teff\ at which fast-rotating stars are no longer present varies with \logL, with cooler fast-rotating stars toward lower \logL, even beyond the TAMS. We argue that this could be a consequence of rotationally induced effects that change the width of the MS.

Based on the distribution of \vsini\ values with \Teff, we argue that the evolution and the larger fraction of the slowly rotating component may be an indicator that massive stars enter the MS with low to mild initial rotations ($v_{\rm ini}$/\vcrit\,\ls0.2). This would be further supported under the assumption that the fast-rotating component mainly emerges from binary interaction.

Using multi-epoch spectra, the associated peak-to-peak \vrad\ variations, and an adequate diagnostic of the intrinsic variability of O-stars and BSGs, we were able to find 176 SB1 systems among 494 stars located within 2500\,pc. We find an average of 39\% of SB1 systems within the MS, which is significantly reduced to 15\% beyond the TAMS, further delineating its position in the HR diagram. We found that the behavior of SB2+ systems with \Teff\ is qualitatively similar to that of SB1 systems.

We compared the location of the theoretical TAMS in the reference sets of evolutionary models of \citet{brott11}, \citet{ekstrom12}, and \citet{choi16b} with its empirical location obtained in this work. Our results suggest a revision of the overshooting parameter for non-rotating models, with higher overshooting values in the first two sets of models and a lower value in the latter set.
We show that the behavior of the FR-TAMS is closer to that displayed by the theoretical TAMS in the models of \citet{ekstrom12}.

The results presented in this work provide strong empirical constraints for improving stellar evolution models. Particularly with respect to the width of the MS in the massive star domain.
Although our results also agree with the possibility that fast-rotating stars may have binary origin, further investigation is required to confirm this hypothesis. Furthermore, stars evolving blueward may also populate the upper part of the HR diagram near the TAMS.
Determining the surface abundances and analyzing the pulsation properties of the stars in our sample will help to find new clues regarding their evolutionary nature and will be addressed in a future work.
Moreover, the fraction of spectroscopic binaries within our sample demands further investigation and will likely add important constraints on the frequency of binary interaction in massive star evolution.


\begin{acknowledgements}

The authors thank the anonymous reviewer for a constructive and positive report that has led to the improvement of the manuscript. The authors also thank Dominic Bowman for providing valuable comments on this manuscript.
AdB, SS-D, and GH acknowledge support from the Spanish Ministry of Science and Innovation and Universities (MICIU) through the Spanish State Research Agency (AEI) through grants PID2021-122397NB-C21, PID2022-136640NB-C22, 10.13039/501100011033, and the Severo Ochoa Program 2020-2023 (CEX2019-000920-S).
Regarding the observing facilities, this research is based on observations made with the Mercator Telescope, operated by the Flemish Community at the Observatorio del Roque de los Muchachos (La Palma, Spain), of the Instituto de Astrof\'isica de Canarias. In particular, obtained with the HERMES spectrograph, which is supported by the Research Foundation - Flanders (FWO), Belgium, the Research Council of KU Leuven, Belgium, the Fonds National de la Recherche Scientifique (F.R.S.-FNRS), Belgium, the Royal Observatory of Belgium, the Observatoire de Genève, Switzerland, and the Thüringer Landessternwarte Tautenburg, Germany.
This work is also based on observations with the Nordic Optical Telescope, owned in collaboration by the University of Turku and Aarhus University, and operated jointly by Aarhus University, the University of Turku and the University of Oslo, representing Denmark, Finland and Norway, the University of Iceland and Stockholm University, at the Observatorio del Roque de los Muchachos, of the Instituto de Astrof\'isica de Canarias.
Additionally this work is based on observations obtained with the FEROS spectrograph attached
to the 2.2\,m MPG/ESO telescope at the La Silla observatory (Chile).
This work has used data from the European Space Agency (ESA) mission {\it Gaia} (\url{https://www.cosmos.esa.int/gaia}), processed by the {\it Gaia} Data Processing and Analysis Consortium (DPAC, \url{https://www.cosmos.esa.int/web/gaia/dpac/consortium}). Funding for the DPAC has been provided by national institutions, in particular, the institutions participating in the {\it Gaia} Multilateral Agreement.

\end{acknowledgements}


\typeout{}
\bibliographystyle{aa} 
\bibliography{biblio} 


\begin{appendix}

\onecolumn 

\section{Further notes on the completeness of the sample}
\label{apen.complet_sptNS}

\begin{figure}[H]
 \centering
    \includegraphics[width=0.46\textwidth]{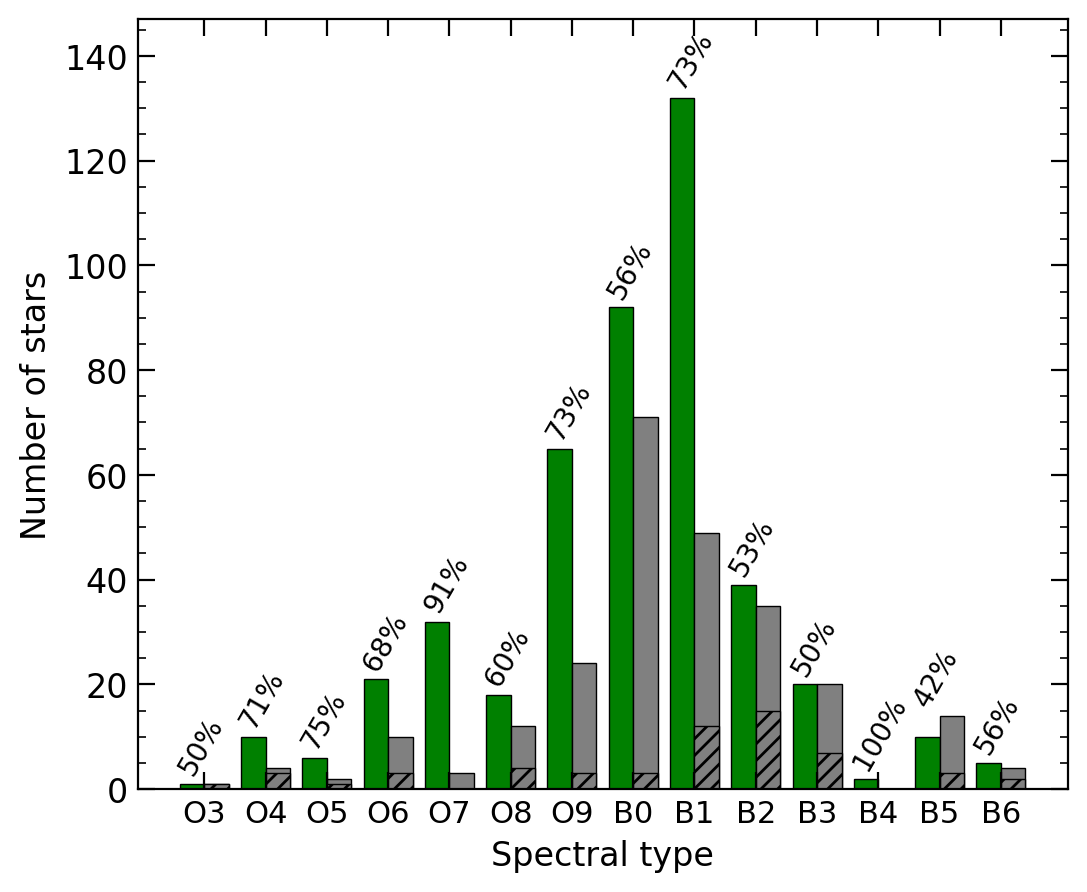}
    \caption{Histogram of the number of stars within 2500\,pc located above the top reddening line of Fig.~\ref{fig:gaia_alsIII} in bins of spectral type. As in that figure, we use green and gray colors to separate analyzed stars in our sample from missing stars listed in the ALS~III catalog (see Sect.~\ref{section:3_tmp} for further details). The dashed pattern within the missing sources corresponds to observed stars that have not been analyzed.}
\label{fig:hist_complet}
\end{figure}

In Sect.~\ref{section:3_tmp} we evaluated the completeness level of the stars analyzed with respect to the ALS~III catalog.
Figure~\ref{fig:hist_complet} extends this by showing the observed and missing stars within 2500\,pc, separated by their spectral type. We do this to further evaluate the potential existence of biases in the distribution of the stars shown in this work that may lead to misinterpretations of the results. 

For most of the spectral types, the percentage of observed stars is higher than the missing ones. Only the B5 spectral type has more missing objects. The B0-type gathers more missing stars than any other group. The associated mean \Teff\ of the observed objects that corresponds to this spectral type is 28\,kK. Interestingly, this corresponds to a temperature close to the TAMS, on the hot side. Having these stars would probably strengthen the position of the TAMS with respect to what already shown. 

The dashed pattern included in the bins of Fig.~\ref{fig:hist_complet} indicates stars for which we have available spectra, but we were unable to provide reliable estimates of the parameters. This fact has not been taken into account in the completeness analysis of Sect.~\ref{section:3_tmp} as we only aim to evaluate the level of completeness of the analyzed stars regardless of whether we have observed them or not. However, we can see that for many of the spectral types, they represent an important fraction of the missing objects. These cases usually correspond to peculiar objects or with problematic features in the spectra (e.g., strong nebular emission) or even stars simply not yet analyzed.
Based on these results, we conclude that the lack of observed or analyzed objects should not affect the interpretations made in this work.

\section{Evidence of disks near critical velocity}
\label{apen.disks_vcrit}

Oe and Be stars that exhibit disks are not present in our sample because of the impossibility of analyzing these objects with conventional 1-D atmospheric models (see Sect.~\ref{section:2_tmp}). Despite this, in \citet{deBurgos24a} some objects exhibiting a small double sub-peak structure in the H$\alpha$ line were analyzed, suggesting the presence of some material around them.
Looking at Fig.~15 of \citet{deBurgos24a}, we observed that the location of these objects in the sHR diagram shifts with \Teff\ and \logLs\ in a similar way as does the drop of fast-rotating stars in the HR diagram\footnote{We note that the presence of a disk might affect the apparent magnitude of these objects and possibly lead to lesser reliable distances. However, we decided to include them.}. 
To further explore this connection, we marked with lime crosses in Fig.~\ref{fig:scatter_vsini_teff} those stars flagged in \citet{deBurgos24a} for having such sub-peak structure. 
Interestingly, we can see that most of them are located close to their \vcrit. As explained in \citet{demink13}, a consequence of stars reaching near-critical rotation is the formation of a near-Keplerian disk, which helps stars to decrease their angular momentum. 
We suggest that our results indicate clear evidence of this phenomenon.


\section{The minimum observed spin rates increasing with luminosity}
\label{apen.min-vsini_logL}

The associated 10\%-percentiles of the top three panels of Fig.~\ref{fig:scatter_vsini_teff}, representative of the low-\vsini\ component (Sect.~\ref{subsection:42_spinrates}), clearly show an almost flat trend, with a slow decay of $\approx$20\kms from the hot- to the cool-end of temperatures in the top two panels. Although the connection between the minimum measured \vsini\ values with the spectroscopic luminosity \citep{holgado22, deBurgos24a} or the evolutionary mass \citep{markova14} has already been shown, our results display for the first time this characteristic with respect to the luminosity. For completeness, the average values of the 10\%-percentiles drop by $\approx$10\kms from the top to the third panel. However, as pointed out in \citet{sundqvist13b} \citep[see also][]{simon-diaz14a}, we may be overestimating our measured values in the low-velocity regime as a consequence of additional turbulent motions present in the star's atmospheres (namely micro- and macroturbulence).


\section{Peak-to-peak detection thresholds for SB1 systems}
\label{apen.rvpp_thresholds}

\begin{figure}[H]
 \centering
    \includegraphics[width=0.46\textwidth]{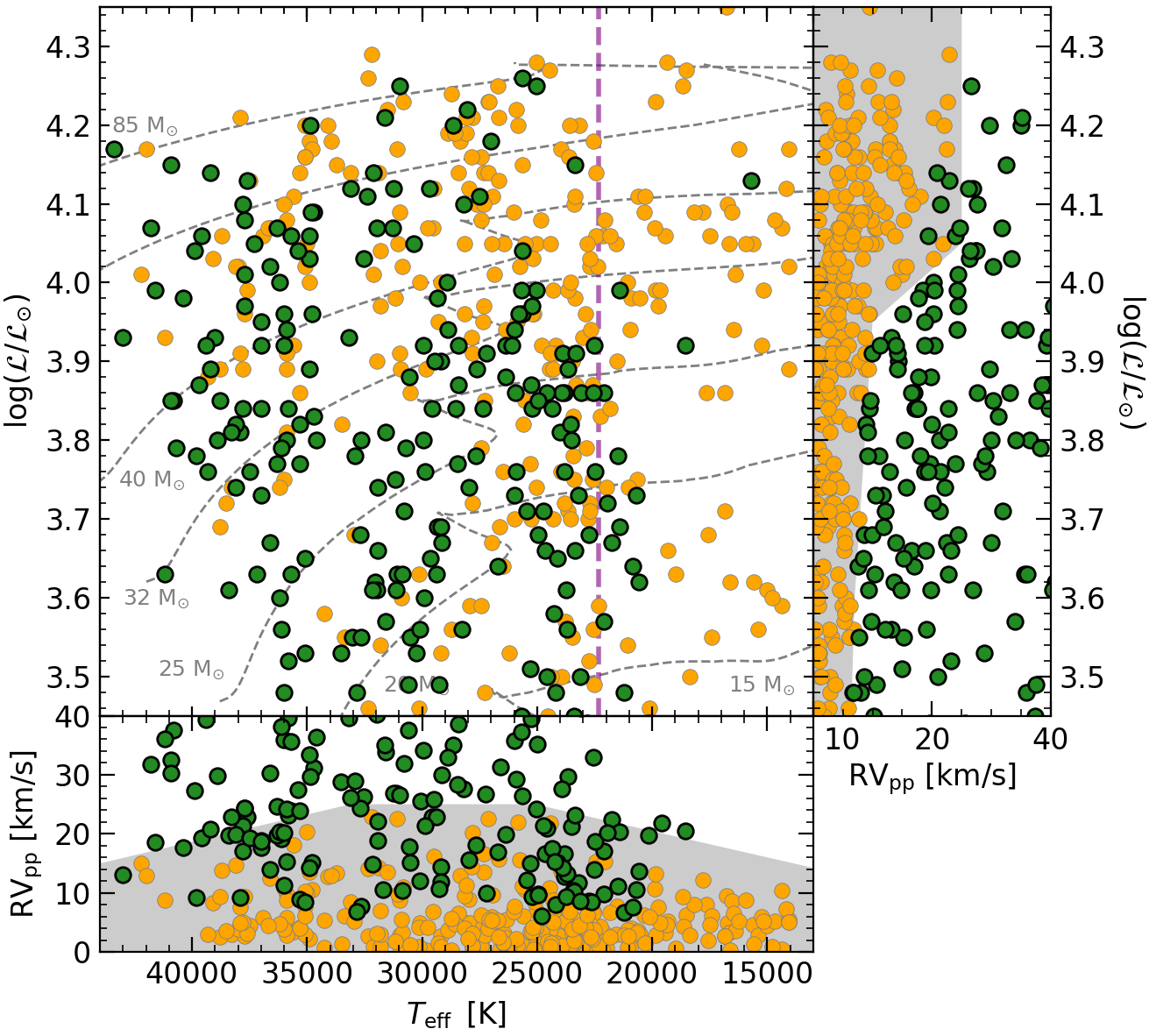}
    \caption{spectroscopic Hertzsprung-Russell diagram of the stars in the sample within 2500\,pc with three or more available multi-epoch spectra. Green circles correspond to systems identified as SB1. The bottom and right subpanels show \dvrad\ against \Teff\ and \logLs, respectively. The gray areas highlight the threshold limits for detection of SB1 systems (see Appendix~\ref{apen.rvpp_thresholds}). The evolutionary tracks correspond to nonrotating Geneva models. The average \Teff\ of our proposed TAMS is shown with a dashed purple line.}
\label{fig:rvpp_thresholds}
\end{figure}

Figure~\ref{fig:rvpp_thresholds} shows the same stars with three or more available multi-epoch spectra and SB1 systems shown in the HR diagram of Fig.~\ref{fig:sb1_hr}, but this time in an sHR diagram. The figure includes two subpanels showing the \dvrad\,--\,\Teff\ and \logLs\,--\,\dvrad\ detection thresholds (gray areas) used in \citet{simon-diaz24a} to separate intrinsic stellar variability from SB1 systems (see Sect.~\ref{subsection:52_binaries}). As explained there, the latter corresponds to those cases in which the \dvrad\ value is above any of the two threshold limits. 

\section{\texorpdfstring{Spin rate properties of stars with \logL\,=\,4.65\,--\,4.3}{Spin rate properties of stars with logL = 4.65-4.3}}
\label{apen.vsini_4.7-4.3}

The bottom panel of Fig.~\ref{fig:scatter_vsini_teff} has some characteristics that are different from the other panels. It is the only panel in which the median values are above the threshold limit of 100\kms, resulting from the smaller fraction of stars in the low-\vsini\ component (see the corresponding panels of Fig.~\ref{fig:vsini_teff}). Although this characteristic is not new \citep[see, e.g.,][]{hunter08, holgado22}, the higher number of fast-rotating stars allows us to better see the ``diagonal" decrease of \vsini\ in the 25\,--\,17\,kK range. 
This panel also shows that most of the stars beyond the TAMS are fast-rotating objects. However, we note the lack of slowly rotating stars in the 21\,--\,15\,kK range, for which we do not find a specific reason, but it could be related to the selection of stars with spectroscopic luminosities with \logLs\,>\,3.5\,dex in \citet{deBurgos24a}. 
The inclusion of additional stars in that \Teff\ range would shift the location of the TAMS toward cooler temperatures, resulting in a lesser number of fast-rotating objects beyond the TAMS.



\end{appendix}

\end{document}